\documentclass[aps, 10pt, prb, twocolumn, amssymb, amsmath, amsfonts, showpacs, superscriptaddress, a4paper]{revtex4-2}
\usepackage{amsmath, amssymb, amsthm}
\usepackage{times}
\usepackage{xcolor}
\usepackage{graphicx}
\usepackage{physics}
\usepackage{comment}
\usepackage[english]{babel}
\usepackage[
  breaklinks,
  colorlinks=true,
  bookmarks=true,
  bookmarksopenlevel=3,
  bookmarksnumbered=true,
  menucolor=blue,
  linkcolor=blue,
  citecolor=blue,
  urlcolor=blue,
  pdfcreator={GhostScript},
  pdfkeywords={},
  pdfstartview=FitH]{hyperref}

\begin{document}

\def\lptms{Universit\'e Paris-Saclay, CNRS, LPTMS, 91405, Orsay, France.}
\def\lps{Universit\'e Paris-Saclay, CNRS, Laboratoire de Physique des Solides, 91405, Orsay, France.}

\title{Two-fluid coexistence and phase separation in a one dimensional model with pair hopping and density interactions}

\author{Lorenzo Gotta}\email{lorenzo.gotta@universite-paris-saclay.fr}\affiliation{\lptms}
\author{Leonardo Mazza}\affiliation{\lptms}
\author{Pascal Simon}\affiliation{\lps}
\author{Guillaume Roux}\affiliation{\lptms}

\date{\today}

\begin{abstract}
We compute the phase diagram of a one-dimensional model of spinless fermions with pair-hopping and nearest-neighbor interaction, first introduced by Ruhman and Altman, using the density-matrix renormalization group combined with various analytical approaches. 
Although the main phases are a Luttinger liquid of fermions and a Luttinger liquid of pairs,
we also find remarkable phases in which only a fraction of the fermions are paired. In such case, two situations arise: either fermions and pairs coexist spatially in a two-fluid mixture, or they are spatially segregated leading to phase separation.
These results are supported by several analytical models that describe in an accurate way various relevant cuts of the phase diagram. 
Last, we identify relevant microscopic observables that capture the presence of these two fluids: while originally introduced in a phenomenological way, they support a wider application of two-fluid models for describing pairing phenomena.
\end{abstract}

\maketitle

\section{Introduction}

Zero-energy Majorana modes characterise several topological superconducting models~\cite{kitaev_unpaired_2001} and play a pivotal role in numerous scientific research fields and quantum-technology applications. 
The appearance of zero-energy Majorana modes in number-conserving condensed-matter models without imposing any mean-field treatment of the interactions, has recently raised important attention~\cite{Fidkowski_2011, Sau_2011, Cheng2011, Kraus_2013, ortiz_many-body_2014, kells_multiparticle_2015, lang_topological_2015, keselman_gapless_2015, iemini_localized_2015, Kells_2015, Kells_2015b, Cheng_2015, Ortiz_2016, iemini_majorana_2017, Zhiyuan_2017, Guther_2017, Leggett_2017,  Leggett_2018, Chen_2018, Zhang_2018, Wang_2018,  Keselman_2018, Yin_2019, Li_2019, Lapa_2020, Lapa_2020b, Wang_2020, Knapp_2020, Bland_2021}.  
Pairing, namely the fact that two fermions bind together and behave as a unique ``molecular'' object, is a key phenomenon to this goal, 
and indeed Majorana modes are expected to appear at the physical boundaries of inhomogeneous systems, in between paired and unpaired fermionic phases~\cite{Ruhman_2015, Ruhman_2017}.
 
Pairing phenomena have been identified in several one-dimensional lattice models of spinless fermions~\cite{mattioli_cluster_2013, dalmonte_cluster_2015, Kane_2017, he_emergent_2019, Gotta_2020}.  
A paradigmatic way of inducing pairing is through density-density interactions, which could be either short ranged or with tails which decay algebraically (see e.g.  Refs.~\cite{zhuravlev_electronic_1997, zhuravlev_breakdown_2000, zhuravlev_one-dimensional_2001, duan_bond-order_2011, hohenadler_interaction-range_2012, li_ground_2019,Ren_2012,Patrick_2019}). Some of these models have a relevance for the description of dipolar fermionic gases or of fermionic Rydberg-dressed states~\cite{Baranov_2012}.  
More recently, a paired phase induced by a gain in kinetic energy has been identified in a model proposed by J.~Ruhman and E.~Altman (which we dub the \textit{Ruhman-Altman} (RA) \textit{model}) featuring the competition between pair hopping and single-particle hopping~\cite{Ruhman_2017}.
Although at a first sight the Hamiltonian is rather abstract, it bears similarities with several electronic models~\cite{Penson1986,Affleck1988,Doniach_1993,Sikkema1995, Aligia_1997, Bouzerar1997,Japaridze_2001, Hirsch_1990}, spin models~\cite{Bariev_1991, Zadnik_2020, Zadnik_2020b} and cold-atom models~\cite{Bilitewski_2016, Chhajlany_2016}.

The fact that a paired phase and an unpaired phase should be separated by a second-order phase transition with emergent critical properties and a central charge $c=3/2$ is crucial for linking pairing to Majorana modes~\cite{Ruhman_2015, Kane_2017}. This prediction has been verified by several numerical analyses~\cite{mattioli_cluster_2013, dalmonte_cluster_2015, he_emergent_2019, Gotta_2020}.  
Yet, in a recent article, we have
shown that this need not always be the case~\cite{Gotta2020}.
Focusing on the RA model, we have demonstrated the appearance of a phase, that we dubbed \textit{coexistence phase}, where a paired and an unpaired phases coexist in the same spinless fermionic chain.
Our result indicates that an unpaired phase and a paired phase can share the same spatial region without hybridizing, raising the question about the reason for which the system did not feature phase separation.

This article expands an earlier discussion~\cite{Gotta2020} on that problem and has a twofold goal.
The first one concerns the study of the effect of a nearest-neighbor interaction, which enhances or decreases pairing depending on its sign.
We show that the coexistence phase is stable with respect to this term and that unambiguous signatures of a
two-fluid coexistence are present for finite and non-perturbative values of the interaction, both in the attractive and in the repulsive case.  
The onset of phase separation is observed only for even larger interaction strengths, and  has different features depending on the sign of the interaction.
When the interaction is attractive, fermions cluster together in a small region of the lattice.
When the interaction is repulsive, the paired and fermionic phases become immiscible and are spatially separated.
Our study demonstrates the thermodynamic stability of the coexistence phase and proves that it is not a unphysical artifact of a fine-tuned model.

Our second goal is to further elaborate on the usefulness of a many-fluid theory in order to describe the competition between phases of paired and of unpaired fermions~\cite{Kane_2017} (the other possible technique being the use of an emergent mode~\cite{Ruhman_2017, he_emergent_2019}). We show that it is possible to give a microscopic interpretation to the fluids, that were originally introduced in a purely  phenomenological way. 
Our results open the path to a wider use of this approach by also pinpointing that a field-theory version of the two-fluid model can describe the entire phase diagram for moderate particle-particle interaction.

The article is organized as follows.
In Sec.~\ref{Sec:Model} we present the model and we map out its zero-temperature phase diagram with a numerical analysis based on the density-matrix renormalization group;
simulations characterising all appearing phases are presented.
In Sec.~\ref{Sec:Immobile} we focus on the first goal and discuss the behavior of the system when unpaired fermions acquire a large mass, and are almost immobile. We show the appearance of a peculiar phase separation, with paired and unpaired fermions occupying different regions of the lattice. This phenomenology provides an excellent starting point for a comparison with the coexistence phase, and we report a description of how one phase evolves into the other one.
In Sec.~\ref{Sec:TwoFluid} we focus on the second goal and present some novel numerical data that further support the use of a many-fluid theory, by introducing microscopic numerical observables that behave as the paired and unpaired fluids. 
Our conclusions are drawn in Sec.~\ref{Sec:Conclusions}; the article is supplemented by six appendices.

\section{The model and the phase diagram}
\label{Sec:Model}
\label{Sec:PhaseDiagram}

We consider a one-dimensional lattice of $L$ sites populated by $N$ spinless fermions so that the lattice filling is $n=N/L$. We introduce the fermionic creation and annihilation operators $\hat c^{(\dagger)}_j$
that satisfy canonical anticommutation relations and define the local density operators  $\hat n_j = \hat c_j^\dagger \hat c_j$.
The Hamiltonian of the RA model~\cite{Ruhman_2017} reads:
\begin{align} 
\hat H=& -t\sum_{j}\hat c^{\dag}_j c_{j+1} + \text{H.c.} 
-t^{\prime} \sum_{j} \hat c^{\dag}_{j+1} \hat c^{\dag}_j \hat c_j \hat c_{j-1}+\text{H.c.} \nonumber \\
&+U_1 \sum_{j}\hat n_{j}\hat n_{j+1}\;. \label{hamiltonian}
\end{align}
The physical meaning of the three terms is the following: a standard particle hopping $t$, a pair hopping $t'$, and a nearest-neighbor density-density interaction $U_1$.
The addition of a simple form of fermionic interaction allows us to investigate the stability of the coexistence phase found in a previous work~\cite{Gotta2020}.

We investigate the zero-temperature phase diagram by means of the density-matrix renormalization group (DMRG) algorithm~\cite{white1992,white1993,schollwock_density-matrix_2005}, which represents a state-of-the-art technique to tackle  one-dimensional many-body quantum systems; we use one implementation based on the ITensor library~\cite{itensor} and one based on the historical approach~\cite{white1992,white1993} with a warmup procedure. 
We have performed numerical simulations with both open boundary conditions (OBC) and periodic boundary conditions (PBC) for a wide range of parameters, keeping up to $m=2600$ states and reaching sizes up to $L=200$ and $L=56$, respectively.

\begin{figure}[t]
\includegraphics[width=\columnwidth]{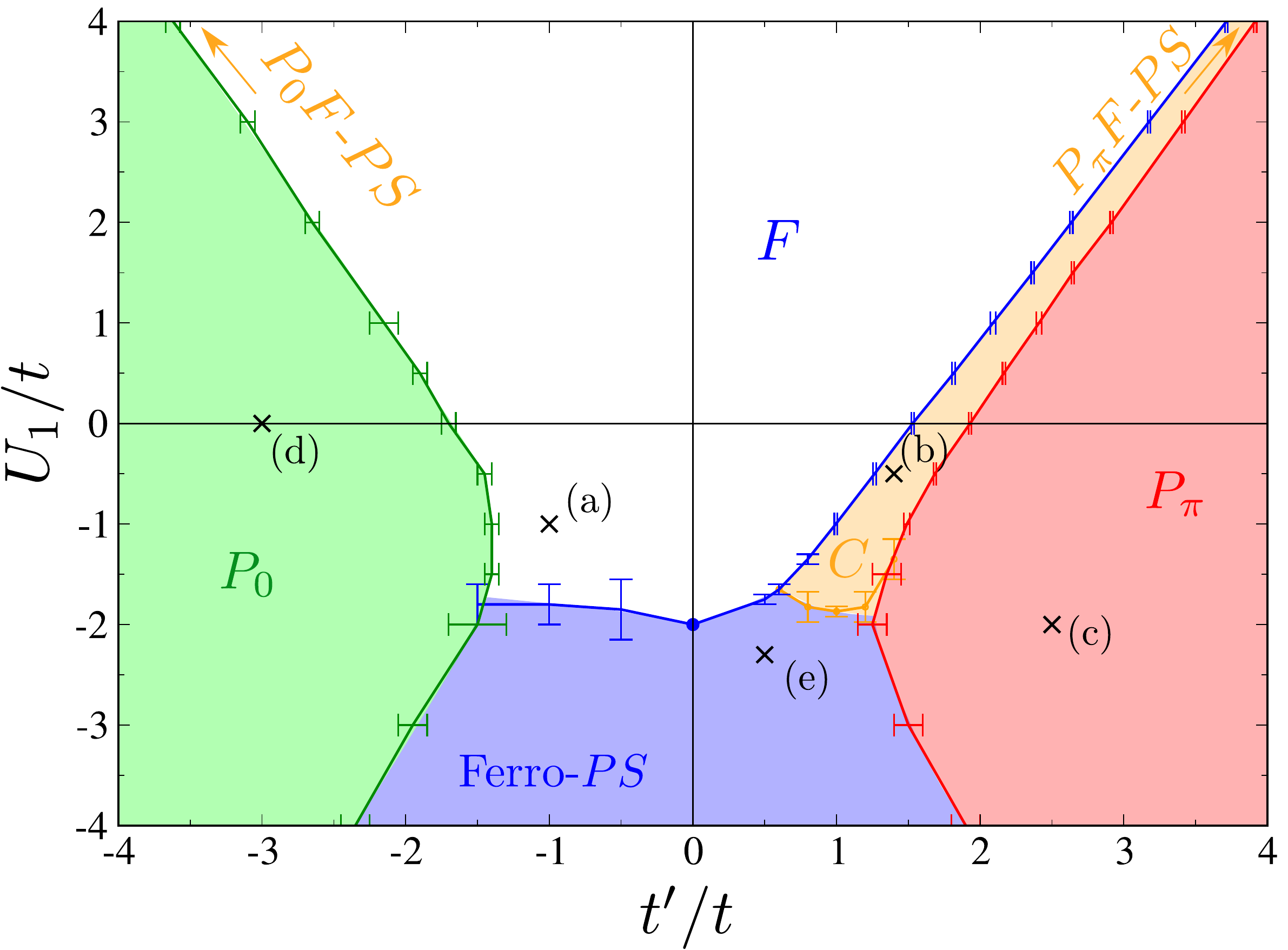}
\caption{Phase diagram of Hamiltonian~\eqref{hamiltonian} for fermionic density $n=1/4$, obtained with DMRG simulations. $F$ is a Fermi Luttinger liquid phase, $P_{0,\pi}$ are paired phases, $C$ is the coexistence phase and $PS$ stands for phase separation, which can be of
different kinds (see text). The crosses and labels (a-e) refer to the various panels of Fig.~\ref{Fig:Correlators}.}
\label{fig:phasediag}
\end{figure}

\begin{figure*}[t]
\includegraphics[width=\textwidth]{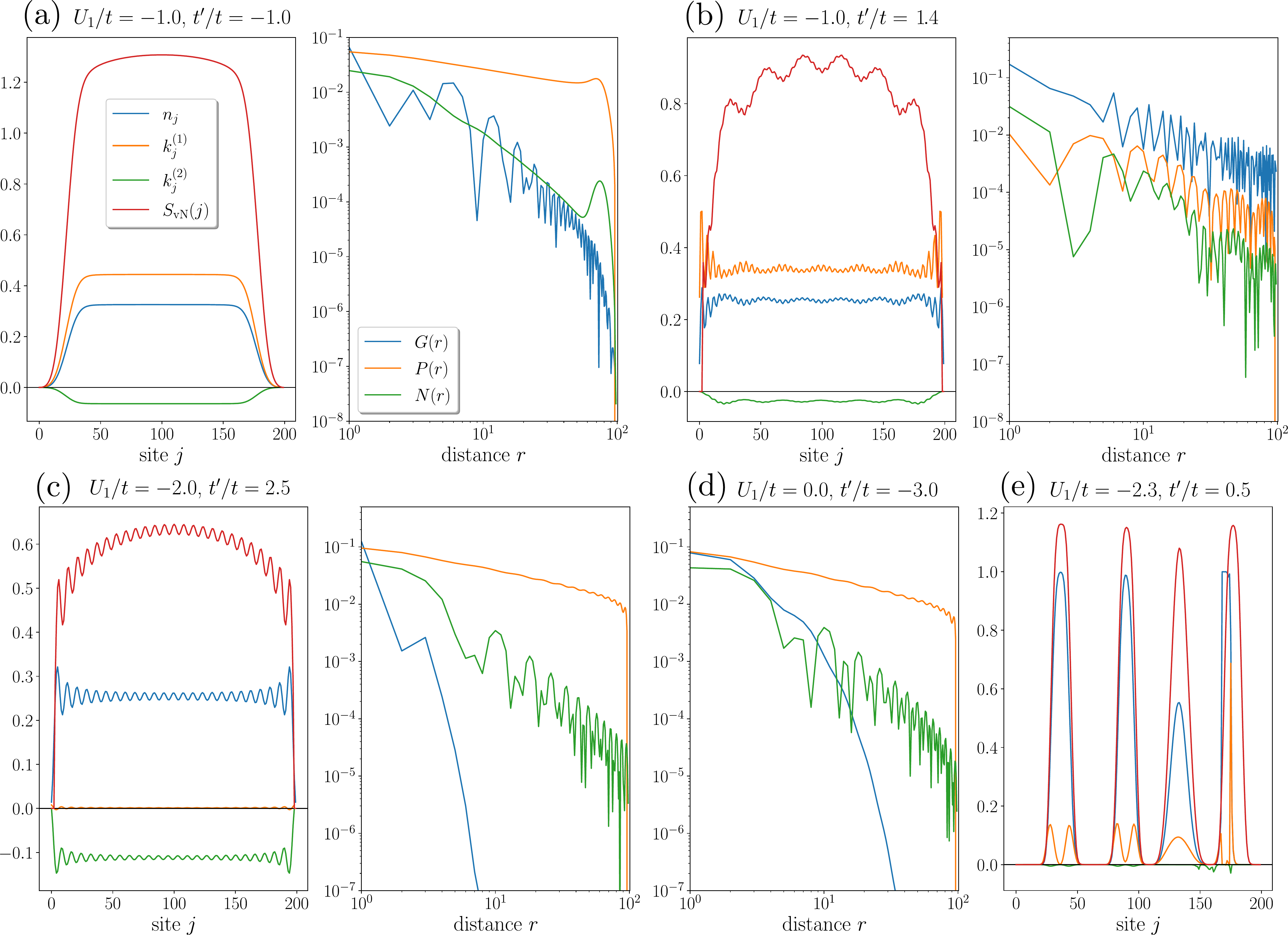}
 \caption{Local observables $n_j$, $k^{(1)}_j $, $k^{(2)}_j$ and $S_{\text{vN}}(j)$, together with absolute values of correlators $G(r)$, $P(r)$ and $N(r)$ for the crosses with labels of Fig.~\ref{fig:phasediag}: (a) F phase (b) C phase (c) P$_\pi$ phase (d) P$_0$ phase (e) Ferro-PS phase.}
 \label{Fig:Correlators}
\end{figure*}

The phase diagram that we obtained for a density $n=1/4$
is presented in Fig.~\ref{fig:phasediag}.
The nature of the phase and the estimated transition lines are based on the behavior of local observables, correlations and entanglement entropy in the system.
We use the local density $n_j=\langle{\hat n_j}\rangle$, the local kinetic energies of particles and pairs defined by
\begin{subequations}
\label{eq:k1_2}
\begin{align}
k^{(1)}_j &= -  \langle \hat c_j^\dagger c_{j+1} + H.c. \rangle,
\\
k^{(2)}_j &= - \langle \hat c_j^\dagger \hat c_{j+1}^\dagger \hat c_{j+1} \hat c_{j+2} + H.c. \rangle;
\end{align}
\end{subequations}
and the entropy profile $S_{\text{vN}}(j)$, which is the entanglement entropy between the two subsystems cut at $(j,j+1)$. The central charge $c$ is obtained by fitting the entanglement entropy (see method described in Ref.~\cite{Gotta2020}).
Regarding correlators, we consider the single-particle Green's function $G(r)=\langle \hat c^{\dag}_{j}\hat c_{j+r}\rangle$, the pair correlator
$P(r)=\langle{ \hat c^{\dag}_i \hat c^{\dag}_{i+1} \hat c_{i+r} \hat c_{i+r+1}}\rangle$ and the density correlator $N(r)= \langle \hat n_{j}\hat n_{j+r}\rangle-\langle \hat n_{j}\rangle\langle\hat n_{j+r}\rangle$, with $j$ taken in the middle of the chain.
These observables are sufficient to discriminate the various phases of Fig.~\ref{fig:phasediag} and their typical behavior in each phase is reported in Fig.~\ref{Fig:Correlators}. 

Looking at Fig.~\ref{fig:phasediag}, we see four main phases corresponding to simple limiting cases. First, the fermionic phase (F) is a standard Luttinger liquid phase with quasi-long-range orders. It is gapless with a central charge $c=1$, and contains the free fermions point $t'=U_1=0$. The dominating fluctuations depend on the Luttinger parameter, which continuously varies across the phase. Qualitatively, positive $U_1$ favors density fluctuations and configuration with non-adjacent fermions. Negative $U_1$ favors neighbouring pairs and eventually drives the system into phase separation. The $t'$ term also favors pairing through a gain in kinetic energy for paired configurations, as readily seen in \eqref{hamiltonian}. Thus, one has regions in the F phase in which the pairing correlations $P(r)$ are the leading ones. Such situation is displayed in Fig.~\ref{Fig:Correlators}(a) where pairing is the leading fluctuations. Yet, we do not call that a paired phase since there is no single-particle gap.

Interestingly, we observe in Fig.~\ref{Fig:Correlators}(a) that observables vanish at the edges of the fluid with OBC. This can be seen as a precursor to phase separation. Indeed, for $t'=0$, the model maps onto the XXZ model, through Jordan-Wigner transformation, which is known to have a transition to a ferromagnetic phase for $U_1=-2t$. For large negative $U_1$, the system naturally forms large domains that become degenerate. Hence, we call this phase Ferro-PS since ferromagnetic domains maps onto fermionic domains. Numerically, this phase separated phase typically displays domains and unequally distributed particles such as in Fig.~\ref{Fig:Correlators}(e), since DMRG is a variational approach that cannot discriminate between the many low-lying states of such phase.
Consequently, the transition line is not easy to investigate resulting in relatively large error bars on Fig.~\ref{fig:phasediag} for this transition. Yet, we observe that it almost does not depend on $t'$.

Focusing  on the effect of $t'$, a large value of $t'$ naturally favors pairs formation, whose nature depends on the sign of $t'$~\cite{Ruhman_2017,Gotta2020}. Pairs condense at wavevector $k=\pi$ for positive $t'$ in the $P_\pi$ phase, and at wavevector $k=0$ for negative $t'$ in the $P_0$ phase (see Sec.~\ref{Sec:FermBoseDOF} for details). These phases have a single-particle gap corresponding to the cost of breaking a neighbouring pair. 
As seen in Fig.~\ref{Fig:Correlators}(c-d), $G(r)$ is exponentially suppressed while pairing fluctuations $P(r)$ are algebraic and dominating. These phases effectively correspond to single-mode Luttinger liquid of pairs with central charge $c=1$.
Pairs, which can be viewed as tightly bound neighbouring fermions, are directly visible on local observables in Fig.~\ref{Fig:Correlators}(c) where each local maximum qualitatively corresponds to a pair ($N/2 = 25$ on total for this system size).

It was early predicted that the transition between P$_0$ and F is direct and second-order~\cite{Ruhman_2017}; the critical point has central-charge $c=3/2$. Indeed, at the phase transition the gapless mode $c=1$ is accompanied by the appearance of an emerging Ising mode with $c=1/2$. This physics has been naturally linked to that of a Kitaev chain and to the emergence of Majorana modes in appropriate situations. Numerically, the transition line displayed in Fig.~\ref{fig:phasediag} is obtained by searching for the maximum of the central charge along fixed-$U_1$ cuts for a $L=200$ size system. We did not systematically check the $c=3/2$ expectation but the findings are compatible with the results of Ref.~\cite{Ruhman_2017}.

\begin{figure}[t]
\centering
\includegraphics[width=\columnwidth]{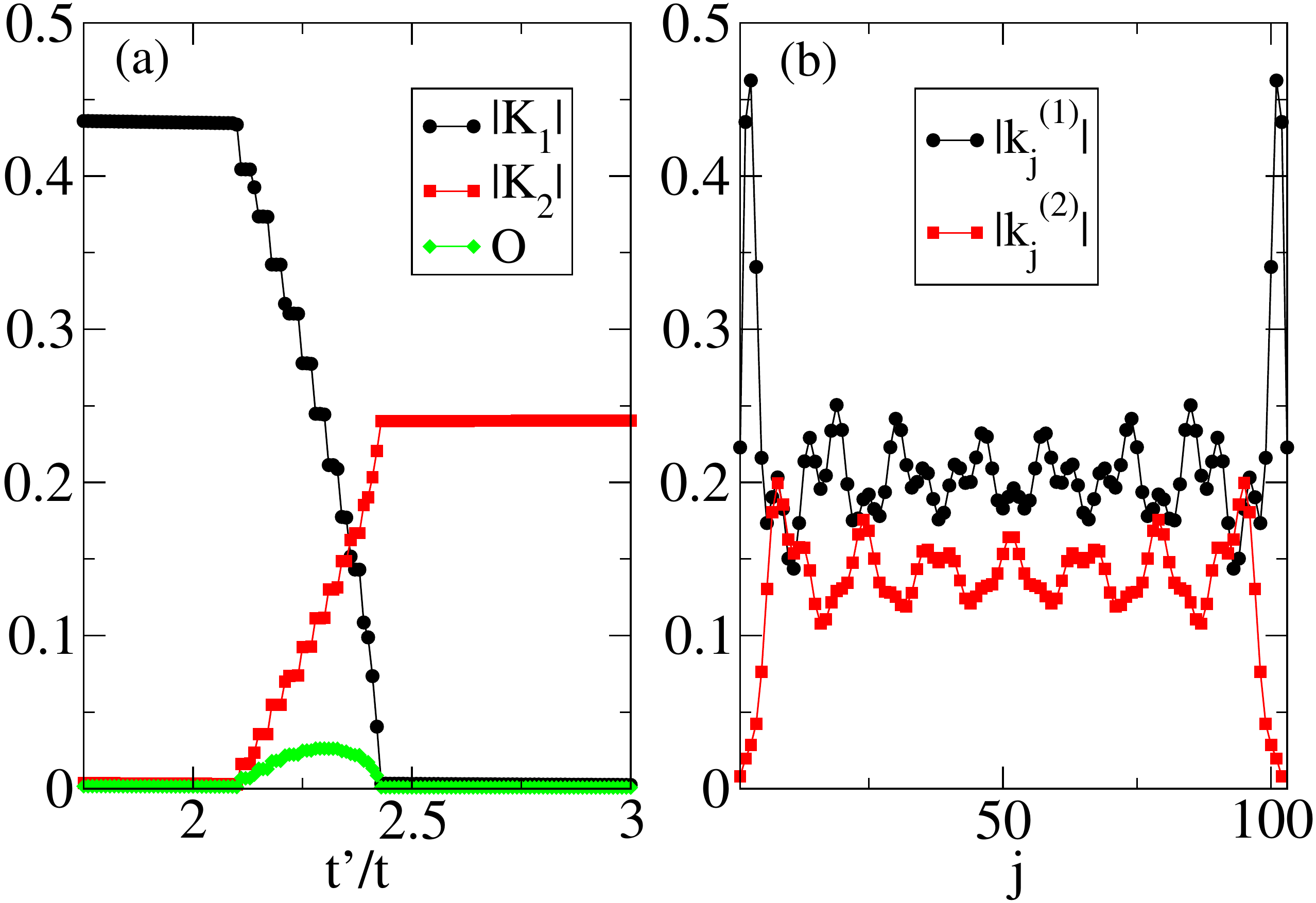}
\caption{Cut along the $U_1/t=1$ line showing the C phase with $L=104$ and OBC. (a) total kinetic energies $K_1$, $K_2$ and overlap $O$ as a function of $t'/t$, see definitions in Eqs.~\eqref{Eq:Kinetic:Energies} and~\eqref{Eq:Overlap}. (b) local particle and pair kinetic energy profiles $k_j^{(1)}$ and $k_j^{(2)}$, for $t^{\prime}/t=2.31$.}
\label{fig:KE_C_phase}
\end{figure}

The most remarkable feature of the phase diagram is the existence of an intervening phase between the F and $P_\pi$ phase, the so-called coexistence phase C found in Ref.~\cite{Gotta2020} on the $U_1=0$ line.
We show that it actually expands to a wide region of the diagram both for attractive and repulsive $U_1$.
The physical picture for the C phase emerging from the local observables of Fig.~\ref{Fig:Correlators}(b), is that part of the fermions are paired (main bumps in the local density) while other lie unpaired in between pairs (smaller oscillations). It is thus a miscible phase of the $P_\pi$ and F phase whose fractions are determined by energy minimization, as we will see later. The two miscible fluids are two intertwined Luttinger liquids that do not hybridize. Consequently, all correlators of Fig.~\ref{Fig:Correlators}(b) capture both contributions from each fluids, with all algebraically decaying fluctuations. 
Note that the behaviour is qualitatively different from that of the F phase: although in both situations they show an algebraic decay, $G(r)$ and $P(r)$ share similar power-law in the C phase. Last, the central charge clearly supports a $c=2$ ``two-fluid" phase~\cite{Gotta2020}.

One characteristic signature of the C phase is that the number of paired fermions continuously evolves from 0 (in the F phase) to $N/2$ (entering the $P_\pi$ phase) across the region. This has a direct consequence in the magnitude of the particle and pair kinetic terms in Eqs.~\eqref{eq:k1_2}. Numerically, we track the total one-particle and pair kinetic energies defined as
\begin{align}
K_1 = \frac{1}{L} \sum_j k^{(1)}_j, \quad 
K_2 =  \frac 1L \sum_j k^{(2)}_j .
\label{Eq:Kinetic:Energies}
\end{align}
to infer with precision the boundaries of the C phase, extending over the interval $ 2.11 < t'/t < 2.43$ in the case $U_1/t=1.0$.
Fig.~\ref{fig:KE_C_phase}(a) shows that one can easily discriminate between the F, C and P$_\pi$ phase from these simple observables, since $K_1$ and $K_2$ are constant in the F and P$_\pi$ phases. 
The little jumps observed signal the creation of one pair and the disappearance of two unpaired fermions, and are thus finite-size effects. As expected, the local profiles $k^{(1)}_j$ and $k^{(2)}_j$ displayed in Fig.~\ref{fig:KE_C_phase}(b) are delocalized over the whole lattice, with opposite maxima, in agreement with the physical picture of the C phase.
Last, we mention that clear signatures of the C phase are also obtained by looking at the density structure factor (see Appendix~\ref{App:Struct:Factor:C} for more details).

\begin{figure}[t]
\centering
\includegraphics[width=\columnwidth]{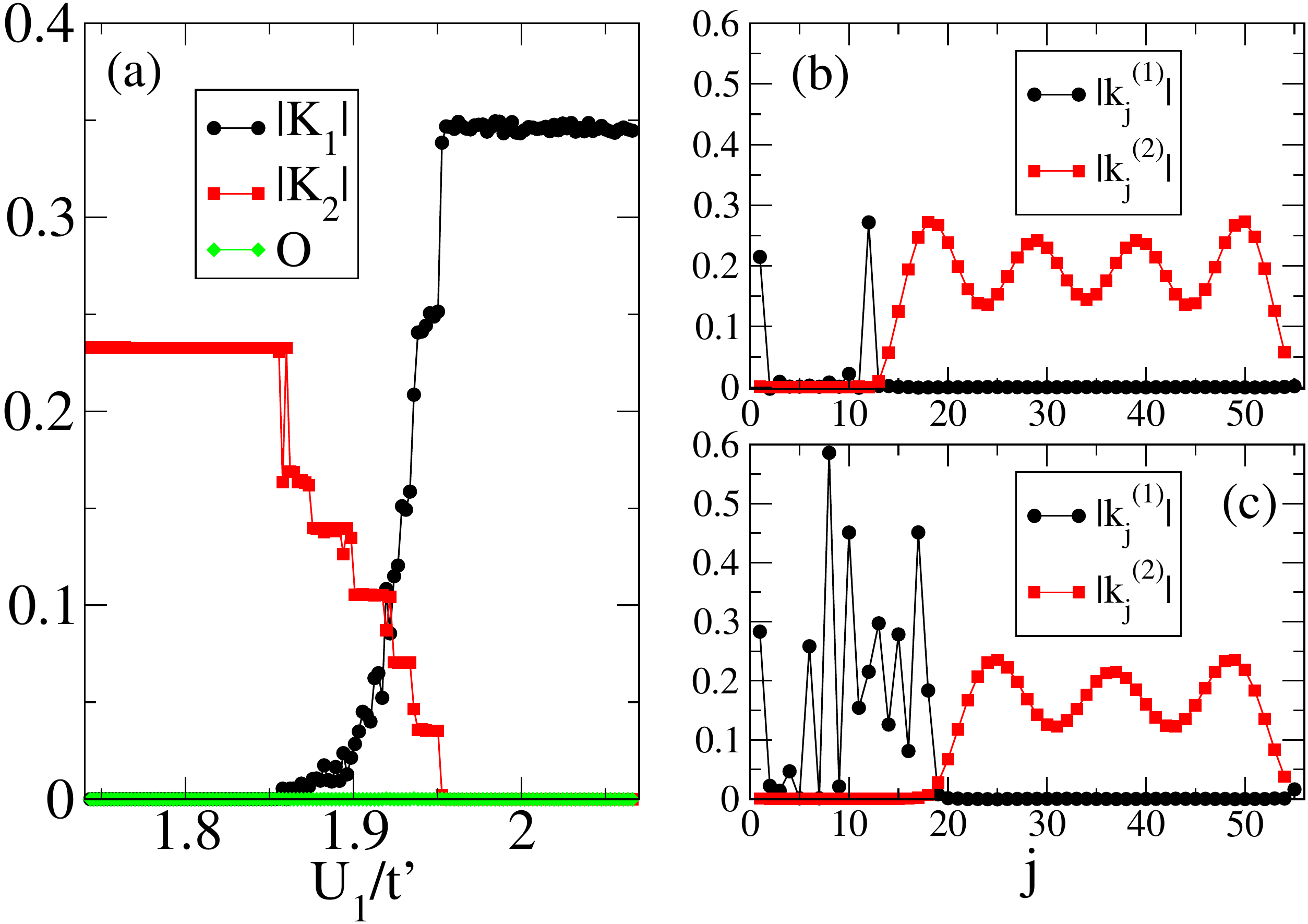}
\caption{Cut showing the P$_\pi$F-PS phase for a chain with $L=56$ and PBC. In this small $t$ limit, we fix $t/t' = \frac{1}{2000} \sqrt{1+(U_1/t')^2} $. 
(a) Kinetic energies $K_1$, $K_2$ and overlap $O$ as a function of $U_1/t'$.
(b) Local kinetic energy profiles $k_j^{(1)}$ and $k_j^{(2)}$ for $U_1/t^{\prime}=1.891646$, indirectly showing a CDW fermionic domain.
(c) Same for $U_1/t^{\prime}=1.912426$, without CDW domain.}
\label{fig:KE_PS_phase}
\end{figure}  

Last, we have denoted two extra phases on Fig.~\ref{fig:phasediag} with arrows: P$_\pi$F-PS and P$_0$F-PS. They occur for a value of $U_1$ that is so large that it cannot be represented on the same diagram for sake of readability. The P$_\pi$F-PS realizes the possibility that, in the presence of paired and unpaired fermions, the two fluids separate spatially into two domains separated by an interface. Thus, we call it a phase separation regime but it is quite different in nature from the Ferro-PS phase. This phase occurs in the low-$t$ limit and is scanned by varying $U_1/t'$ for a small $t$. The behavior of the particle and pair kinetic energies along such cut is displayed on Fig.~\ref{fig:KE_PS_phase}(a). We see a qualitatively similar behavior as in Fig.~\ref{fig:KE_C_phase}(a), with the difference that the F phase is now on the right hand side and the P$_\pi$ on the left. These observables simply show that there is an intervening phase that comprises both pairs and unpaired fermions, whose numbers are continuously varying. In order to distinguish this phase from the C phase, one has to look at local observables, as reported on Fig.~\ref{fig:KE_PS_phase}(b-c). There, we clearly observe two domains separated by domain walls. Numerically, this state is best reached for PBC since OBC add two egdes to the chain that interfere with domain formation. Furthermore, we observe two different behaviors for the fermionic domains: in Fig.~\ref{fig:KE_PS_phase}(b), the local kinetic energy $k^{(1)}_j$ vanishes while it fluctuates significantly in Fig.~\ref{fig:KE_PS_phase}(c). The reason is that in this large $U_1/t$ limit, fermions are strongly repulsive and the highest density domain that can be formed is a charge-density wave (CDW) state $\ket{\bullet\circ\bullet\circ\bullet\circ\cdots}$ that has almost no residual kinetic energy. Local density profile (not shown) is in full agreement with this interpretation for Fig.~\ref{fig:KE_PS_phase}(b).

We understand that total kinetic energies $K_{1,2}$ are not space-resolved and do not help distinguish the C phase from the P$_\pi$F-PS phase. For this reason, we introduce the "overlap" 
\begin{equation}
 O = \frac 1L \sum_j |k_j^{(1)}| \times |k^{(2)}_j| \;.
\label{Eq:Overlap}
\end{equation}
It captures the spatial correlations between the two kinetic energy profiles. Intuitively, in the C phase it should be non-zero because pairs and unpaired fermions delocalize over the same regions. On the contrary, it should be zero in the P$_\pi$F-PS. Fig.~\ref{fig:KE_C_phase}(a) and \ref{fig:KE_PS_phase}(a) show that, across each intervening phase, the values of $O$ are perfectly consistent with this interpretation. In short, following the behavior of
both $K_{1,2}$ and $O$ allows us to clearly discriminate between the four phases F, P$_\pi$, C and P$_\pi$F-PS. Last, we mention that for negative $t'$, large-$U_1$ DMRG calculations also demonstrate the existence of a P$_0$F-PS intervening phase, which shares the same phenomenology as the P$_\pi$F-PS phase, but with P$_0$ pairs instead.

The rest of the manuscript provides more details and an expanded discussion about the main features of this phase diagram.

\section{Pairs-fermions phase separation vs. coexistence phase} \label{Sec:Immobile}

The goal of this section is to discuss the P$_\pi$F-PS phase, and to characterise its differences with respect to the C phase.
We first consider the region where fermions have little delocalisation tendency and thus are almost immobile, $t \ll t'$, $U_1$. 
We focus on and describe extensively the P$_\pi$F-PS phase, which appears for $1.86 < U_1/t' < 2$ for $t=0$. By performing simulations with non-zero values of $t$, we are able to identify some aspects of the crossover to the C phase.
Note that the results of this section have an interest that goes beyond this work because the $t=0$ limit is relevant for discussing flat-band models, where indeed single-particle hopping is zero or negligible, and density-assisted hopping competes with density-density interaction, see for instance Refs.~[\onlinecite{Tovmasyan_2013, Takayoshi_2013, Junemann_2017}]. We expect that the elementary interpretation that we develop here could shed light also on these physical systems.

\subsection{Overview on the limit $t \ll t', U_1$}

\begin{figure}[b]
\includegraphics[width=\columnwidth]{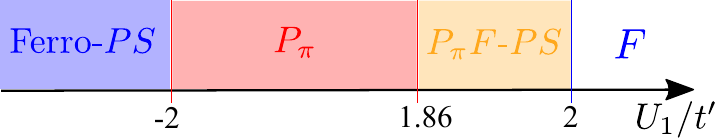}
\caption{Phases of the system for $t \ll t', U_1$ as a function of $U_1/t'$.} \label{Fig:Phases:Smallt}
\end{figure}

Looking at the phase diagram reported in Fig.~\ref{fig:phasediag}, the limit of immobile fermions corresponds to a study along a circumference at large distance from the centre: $U_1/t'$ determines the angle along which one takes the limit $t \to 0^+$.
Varying $U_1/t'$ with positive $t'$, one encounters four different phases along such parametrisation, as summarized in Fig.~\ref{Fig:Phases:Smallt}. 
For weak interactions, $-2<U_1/t'\lesssim1.86$, the system is in a paired phase P$_\pi$.
For strong attractive interactions, $U_1/t'<-2$, the system enters the Ferro-PS phase.
For strong repulsive interactions, $1.86 \lesssim U_1/t'<2$, the system first enters the P$_\pi$F-PS phase.
For more repulsive interactions $U_1/t'>2$, all pairs are broken and the system enters the F phase, that is adiabatically connected to the non-interacting limit.

In the following three subsections, we present details on each of these phases and discuss several effective models that capture the main features of these parts of the phase diagram.
We then study how the P$_\pi$F-PS phase makes a transition to the C phase.

\subsection{The paired phase and the ferromagnetic phase separation}

\subsubsection{Reminder on the case $U_1=0$} \label{paired_subspace}

It has already been shown that the Hamiltonian in Eq.~\eqref{hamiltonian} at $t=U_1=0$ can be diagonalised exactly by mapping the model to an effective XX spin chain~\cite{Chhaijlany_2016,Zadnik_2020, Gotta_2020} and can be interpreted as a paired phase P$_\pi$. We now briefly review the argument.  
We first introduce the subspace $\mathcal{H}_{P}$, where all fermions are paired, which is defined as follows. We consider all Fock (product) states that span the Hilbert space of the one-dimensional setup and retain only those where fermions form clusters of even length. 
For example, the state $\ket{\psi_1}=\ket{\bullet\bullet\circ\bullet\bullet\bullet\bullet\circ\circ}$ belongs to $\mathcal{H}_P$, whereas the state $\ket{\psi_2}=\ket{\bullet\bullet\circ\bullet\circ\circ\bullet\circ\circ}$ does not.
Since the pair-hopping term enhances the kinetic energy of pairs, we assume that the ground state lies in the subspace $\mathcal{H}_{P}$; we retain only the states with $N$ fermions, and thus with $N_b=N/2$ nearest-neighbour pairs. 
The Hilbert space $\mathcal H_P$ can be mapped onto the Hilbert space of a spin-1/2 chain of length $L_b = L-N_b$ with  Pauli operators $\hat \sigma_{j}^{\alpha}$, with $\alpha = x,y,z$. The mapping is defined by the following local rules: $\ket{\bullet\bullet} \rightarrow \ket{\uparrow}$, $\ket{\circ}\rightarrow \ket{\downarrow}$;
thus, a spin up stands for a pair while a spin down stands for an empty site. For example, the state $\ket{\psi_1}$ introduced above gets mapped to the spin state $\ket{\uparrow\downarrow\uparrow\uparrow\downarrow\downarrow}$.
According to this mapping, the $N_b$ term that appears in the definition of $L_b$ is understood as an excluded volume.

The action of the Hamiltonian in Eq.~\eqref{hamiltonian} over the subspace $\mathcal{H}_{P}$ is unitarily equivalent to that of an effective XX spin-1/2 Hamiltonian $\hat{H}_{\rm eff} = t^{\prime}\sum_{j=1}^{L_b}\left[\sigma_{j}^{+}\sigma_{j+1}^{-}+H.c.\right]$ that is diagonalized  by means of a Jordan-Wigner transformation. Written in terms of fermionic Fourier modes, we have $\hat H_{\rm eff}=\sum_{k} \varepsilon_{p} (k) \,\hat n_{k}$, with the pair band dispersion relation $\varepsilon_{p}(k) = 2 t' \cos(k)$. 
The ground-state energy per site takes the form
\begin{equation} \label{XX_GS_energy}
e_{{\rm eff}}=\frac{1}{L}\!\!\!\sum_{|k|< \pi \frac{N_b}{L_b}}\!\!\varepsilon_{p}(k)= -\frac{ 2|t^{\prime}|}{\pi}\left(1-\frac{n}{2}\right)\sin\left( \frac{\pi n}{2-n}\right),
\end{equation} 
where we use $n=2N_b/L$. 
The obtained result for the ground state energy does not depend on the sign of $t^{\prime}$. This is understood from the fact that the unitary transformation $\hat{U}=e^{i\frac{\pi}{2}\sum_j j\hat{n}_j}$, which implements the local transformation rule $c_{j}\rightarrow e^{-i\frac{\pi}{2}j} c_{j}$, connects the two Hamiltonian with opposite signs through the relation $\hat{U}\hat{H}(t=0,t^{\prime},U_1)\hat{U}^{\dag}=\hat{H}(t=0,-t^{\prime},U_1)$.

\subsubsection{Pairing and ferromagnetic phase separation for $U_1 \neq 0$}

As soon as one considers nearest-neighbour interactions, keeping $t=0$, the problem becomes non-trivial and is not amenable to a direct exact diagonalization of the Hamiltonian. As a first approximation, we assume that the ground state of the system lies in $\mathcal{H}_{P}$ and write the effective Hamiltonian restricted to this subspace, which takes the form of an XXZ spin model written in terms of the $\hat \sigma_j^{\alpha}$ Pauli matrices:
\begin{align} \label{XXZ_hamiltonian}
\hat H=&\,t^{\prime}\sum_{j=1}^{L_b}\left[\hat \sigma_{j}^{+}\hat \sigma_{j+1}^{-}+H.c.\right]+\nonumber \\
&+\frac{U_1}{4}\sum_{j=1}^{L_b}\left(1+\hat \sigma_{j}^{z}\right)\left(1+\hat \sigma_{j+1}^{z}\right)+U_1\frac{N}{2}.
\end{align}
The assumption that the ground state lies in $\mathcal H_P$ will certainly be even more valid when $U_{1}<0$, since attractive interactions further enhance pairing. 
The Hamiltonian in Eq.~\eqref{XXZ_hamiltonian} 
thus represents a good model for studying the case of attractive interactions. In the spin language, the XXZ model displays phase separation to a ferromagnetic phase that here corresponds to $U_1/t^{\prime}=-2$. Thus, this model predicts a transition from the paired phase $P_\pi$ to the Ferro-PS at $U_1/t^{\prime}=-2$.
This transition extends up to lower values of $t^{\prime}/t, U_1/t$ as the transition line separating the P$_\pi$ phase from the Ferro-PS phase in Fig.~\ref{fig:phasediag}.~

We now focus on the more interesting case of repulsive $U_1$.
Here we expect that the actual ground state will not only contain pairs, but also unpaired fermions.
Simple energetic arguments allow us to estimate the transition point to the F phase. From Eq.~\eqref{XXZ_hamiltonian}, we see that the cost of breaking a pair is given by $U_1$. On the other hand, the kinetic term creates a Fermi sea of pairs with energies ranging from the bottom of the band, $- 2 t'$, to the Fermi energy, $-2 t' \cos \left(\pi n /(2-n) \right) $ which for $n = 1/4$ is approximately $\sim -1.8 t'$. 
Considering that one can place unpaired fermions at distance larger than one at zero energetic cost, we estimate that the F phase should appear for $U_1/t'>2$ and that the paired phase should not be destroyed for $U_1/t'\lesssim 1.8$.
We tested this scenario with a numerical simulation of the model for $t=0$. In Fig.~\ref{fig:U1_pos}(a) we plot the ground-state energy as a function of $U_1/t'$ and we observe the appearance of a zero-energy ground state for $U_{1}/t^{\prime} \sim 2$. 
The plot shows that our effective model in Eq.~\eqref{XXZ_hamiltonian} describes in a very good way the exact numerical results for $t=0$ and $U_1/t' < 2$.

\subsection{The fermionic phase}

An inspection of the density profile of the ground state obtained for $U_1/t'>2$ shows that it is characterised by the presence of unpaired fermions at a distance larger than one (not shown); it is easy to show that all these states are zero-energy eigenstates of the Hamiltonian and span a subspace of fully unpaired fermions that we dub $\mathcal H_S$.
The Fock states that span $\mathcal H_S$ are efficiently described by observing that they comprise two building blocks, $\ket{\bullet\circ}$ and  $\ket{\circ}$, repeated and alternated in different order. 

\begin{figure}[t]
\centering
\includegraphics[width=\columnwidth]{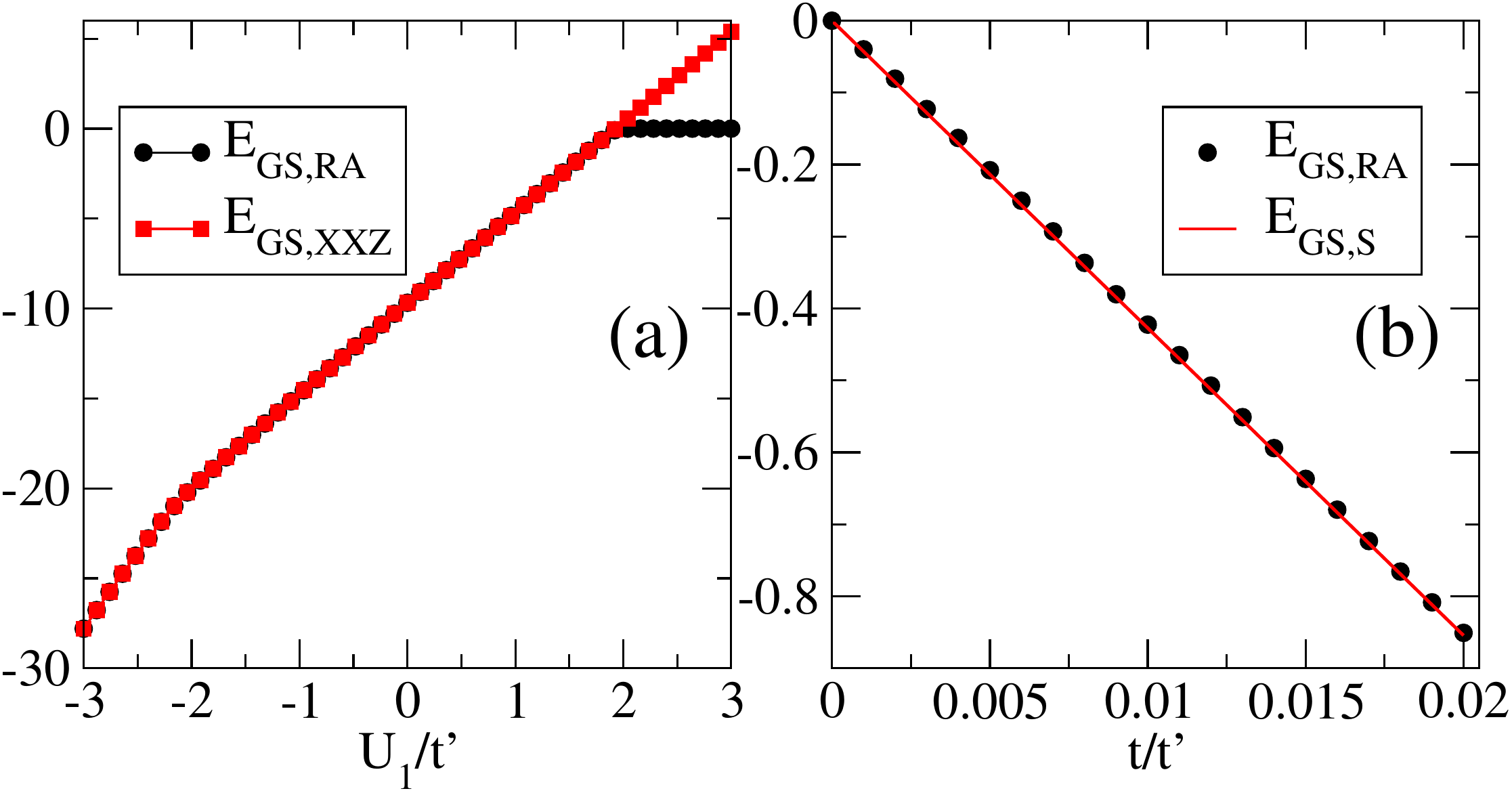}
\caption{Comparison of the ground-state energy $E_{\text{GS,RA}}$ of Hamiltonian~\eqref{hamiltonian} with that of the effective models.
(a) for $t=0$, with $L=40$, $N=10$ and PBC for Hamiltonian~\eqref{hamiltonian}, with $L=35$, $N=5$ and PBC for Hamiltonian~\eqref{XXZ_hamiltonian}. 
(b) for small $t/t'$ with $L=104$, $N=26$, $U_1/t' = 4$ and OBC for Hamiltonian~\eqref{hamiltonian}. The line $E_{\text{GS,S}}$ is the ground state energy of Hamiltonian~\eqref{perturbative_hamiltonian} with the same parameters.}
\label{fig:U1_pos}
\end{figure}

The subspace $\mathcal H_S$ is expected to be adiabatically connected to the F phase once the single-particle hopping is reintroduced in the model.
In order to show it, we consider the limit $t \ll t', U_1$, and we 
derive an effective Hamiltonian restricted to $\mathcal H_S$ using a standard perturbative approach.
We take as unperturbed Hamiltonian the $t=0$ limit of model~\eqref{hamiltonian} and take the term proportional to $t$ as the small perturbation.
Defining $\hat P_S$ the projector onto $\mathcal H_S$, the effective Hamiltonian to first-order in $t$ reads  $\hat H_{S}=\hat P_S \left( -t \sum_j \hat c_j^\dagger \hat c_{j+1}+ H.c.\right) \hat P_S$.
This is the Hamiltonian of mobile fermions which never get to a distance smaller than two.

We now show that $H_{S}$ can be mapped back to spin-1/2 model defined over a spin chain of length $L-N$ with Pauli operators $\hat \Sigma^\alpha_j$, with $\alpha = x,y,z$.
To do so, we use the following rules:
$\ket{\bullet \circ} \to \ket{\uparrow}$ and  $\ket{\circ} \to\ket{\downarrow}$. In particular, we work with $N$ spin-up states.
The single-particle hopping implements a simple exchange dynamics, leading to the following XX model
\begin{equation} \label{perturbative_hamiltonian}
\hat H_{S}=-t\sum_{j=1}^{L-N}\left[\hat \Sigma_{j}^{+} \hat \Sigma_{j+1}^{-}+H.c.\right].
\end{equation}
Again, the ground state energy is evaluated by mapping it to a fermionic problem with a Jordan-Wigner transformation. The energy density in the thermodynamic limit takes the form:
\begin{equation} \label{gs_energy_perturbation}
\epsilon_{GS,S} 
=-\frac{2t(1-n)}{\pi}\sin\left(\frac{\pi n}{1-n}\right).
\end{equation}
A comparison between a finite size and OBC version of formula~\eqref{gs_energy_perturbation} and the corresponding exact DMRG simulation for $t/t'<0.02$ and $U_1/t^{\prime}=4$ is presented in Fig~\ref{fig:U1_pos}(b). 
The result displays a remarkable agreement.

\subsection{Pairs-fermion phase separation}
\label{Sec:PpiFPS}

The numerical study of the model for $t=0$ ($U_1 \sim t'$) is complicated by the existence of conserved quantities. Indeed,
the total numbers of fermions on even and odd sites, $\hat N_e = \sum_j \hat n_{2j}$ and $N_o = \sum_j \hat n_{2j+1}$ respectively, commute with the Hamiltonian. During the matrix product state variational optimization, the values of these conserved quantities will be determined by the initial state. If the latter has not the same numbers as the ground-state, the ground state is not reached. In order to tackle the $t\to 0$ limit, we allow for a small $t \neq 0$ in the Hamiltonian to help the variational state tunnel towards the ground-state. The obtained results are then continously connected to the analytical intepretation of the $t=0$ limit.

In order to study the phase diagram in this limit, we use the following parametrization of the plane of Fig.~\ref{fig:phasediag}: we fix a large radius  $r=\sqrt{\left(\frac{U_1}{t}\right)^{2}+\left(\frac{t^{\prime}}{t}\right)^{2}}$ and vary the angle $\theta = \arctan(U_1/t')$. 
As already anticipated,
the results presented in Fig.~\ref{fig:KE_PS_phase} for $r=2000$ show clear evidence of the existence of an intermediate phase separation regime between the F phase and the P$_\pi$ phase, the P$_\pi$F-PS phase. 
In order to better interpret it, we introduce several effective models that reproduce the ground-state phenomenology found numerically.

\subsubsection{An effective model for $t=0$}

In order to understand this transition region, we formulate an effective model that describes the three main phases that are encountered at $t=0$ namely, the paired phase P$_\pi$, the unpaired F phase and the phase separation P$_\pi$F-PS regime. As it will become clear later, setting $t=0$ considerably simplifies the development of an analytical model, which still describes the numerical data obtained for finite $r$. Indeed, letting $t\rightarrow 0$ is equivalent to let $r\rightarrow\infty$, thus leaving $\theta$ as the only free parameter of the problem.   

\begin{widetext}
The model is based on an ansatz for the energy of generic phase-separated configurations of $N_f$ unpaired fermions and $\frac{N-N_f}{2}$ pairs. We characterize a generic variational configuration with $N_f$ unpaired fermions as follows. On one side, unpaired fermions are immobile, since $t=0$, and form a zero energy CDW domain of length $2N_f$ (with unit cell $\ket{\bullet\circ}$). On the other side, pairs delocalize on the rest of the lattice, a domain of length $L-2N_f$. Their kinetic contribution to the energy density of the configuration is derived analogously to Eq.~\eqref{XX_GS_energy}, provided that $N_b$ equals the number of pairs in the given configuration, namely $\frac{N-N_f}{2}$, and $L_b$ takes the form $L-2N_f-\frac{N-N_f}{2}$, as it equals the size of the lattice region available to pairs, $L-2N_f$, minus the number of pairs, $\frac{N-N_f}{2}$. Finally, the $U_1$ interaction energy density contribution is taken into account in the low density limit by only considering the potential energy density cost $U_1\frac{N-N_f}{2L}$ associated to the formation of $\frac{N-N_f}{2}$ pairs.
After introducing the unpaired fermionic density $n_f=N_f/L$, the ansatz for the ground-state energy density in units of $t'$ as a function of $n_f$ reads
\begin{equation}
\mathcal E(n_f , \theta) = 
-\frac{2}{\pi}\left(1-2n_{f}-\frac{n-n_{f}}{2}\right)\sin\left[\pi\frac{n-n_f}{2\left(1-2n_f-\frac{n-n_f}{2}\right)}\right]+
\frac{n-n_f}{2}\tan\theta,
\label{t_0_PS_model}
\end{equation}
where the relation $U_1/t^{\prime}=\tan\theta$ has been used. 
\end{widetext}

\begin{figure}[b]
\centering
\includegraphics[width=\columnwidth,clip]{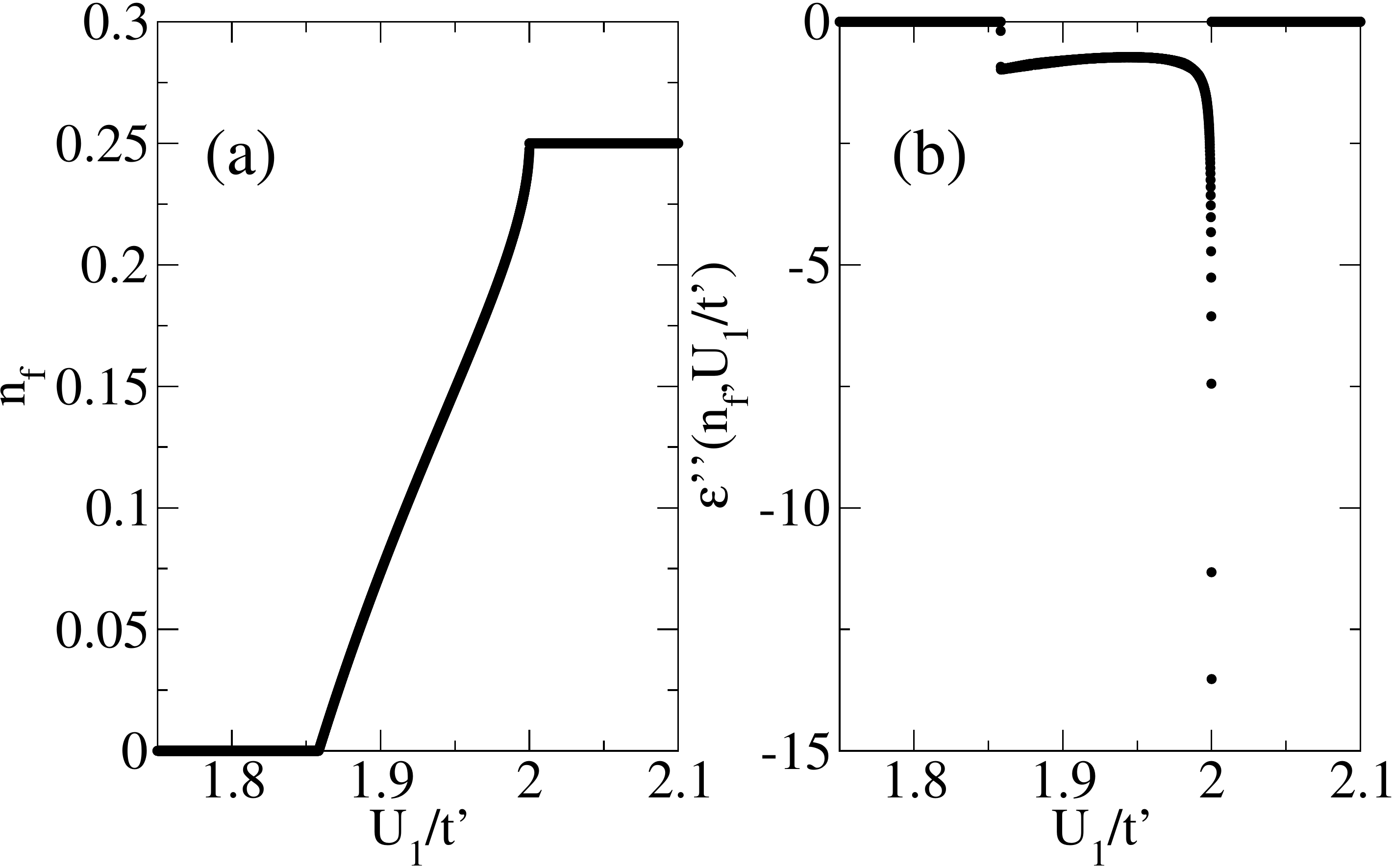}
\caption{(a) Optimal fermionic density $n_f(\theta)$ as a function of $\tan\theta=U_1/t^{\prime}$ according to the effective phase separation model~\eqref{t_0_PS_model}. (b) Second derivative of the energy density $\mathcal E(n_f(\theta), \theta)$ in Eq.~\eqref{t_0_PS_model} as a function of $\tan\theta=U_1/t^{\prime}$.}
\label{fig:PS_density}
\end{figure}

For each $\theta$, we find numerically the optimal value of $n_f$ that minimizes $\mathcal E$. The behavior of the resulting $n_f$ as a function of $U_1/t'$ is shown in Fig.~\ref{fig:PS_density}(a).
For $U_1/t' \lesssim 1.86$, the ground state is fully paired and thus $n_f = 0$. When $n_f=n$, pairing is energetically unfavorable because of the strong nearest-neighbour repulsion.
Thus, the system occupies one of the zero-energy configurations consisting of isolated localized fermions. 
An intermediate value $0<n_f<n$ signals the  onset of phase separation: the system is partitioned into a region of immobile fermions and a region with a liquid of pairs.
We identify this behaviour with the previously-introduced P$_\pi$F-PS phase.

Remarkably, the simple structure of Eq.~\eqref{t_0_PS_model} allows us to obtain analytical results (i) on the position of the critical points, and (ii) on the singular behaviour of the ground state energy-density close to the critical points. Specifically, the P$_\pi\to$P$_\pi$F-PS critical point is given by
\begin{equation}
\left(\frac{U_1}{t^{\prime}}\right)_{c1}=\frac{6}{\pi}\sin\left(\frac{\pi n}{2-n}\right)-\frac{4(2n-1)}{2-n}\cos\left(\frac{\pi n}{2-n}\right)\;.
\end{equation}
For  $n=0.25$, this formula gives $\left(\frac{U_1}{t^{\prime}}\right)_{c1}\approx 1.858$.
Similarly, the critical point P$_\pi$F-PS$\to$F is located at
$\left(\frac{U_1}{t^{\prime}}\right)_{c2}=2$.
Both values match very well with those obtained numerically in Fig.~\ref{fig:KE_PS_phase}(a).

We also derive analytically the singular behaviour of the energy density at the critical points: the second derivative of the energy density with respect to $U_1/t^{\prime}$ shows a finite jump discontinuity for the P$_{0,\pi}\to$P$_\pi$F-PS transition at $(U_1/t^{\prime})_{c1}$ and a square root singularity for the P$_\pi$F-PS$\to$F transition at $(U_1/t^{\prime})_{c2}$. The overall behavior of this second derivative as a function of $U_1/t^{\prime}$ is presented in Fig.~\ref{fig:PS_density}(b).  

\subsubsection{An effective model for $t \neq 0$}

We now test the robustness of the P$_\pi$F-PS phase when $t \neq 0$. The addition of single-particle hopping allows the system to lower its energy through delocalization of unpaired fermions. We generalize the model in Eq.~\eqref{t_0_PS_model} to the case of finite $t$ by introducing an energy gain due to the fermionic delocalization.

\begin{widetext}
Since we do not know a priori the size of the unpaired fermions domain, we introduce an effective fermionic length $l_f=L_f/L$ as a variable. With this notation, the ansatz for the ground-state energy density (in units of $rt$) reads:
\begin{equation} \label{PS_model_general}
\mathcal E_2( n_f, l_f, r, \theta)=\begin{cases}
-\frac{\cos\theta}{\pi}(2-n)\sin\left(\frac{\pi n}{2-n}\right)+\sin\theta\frac{n}{2},~~~~ \text{if}~ (n_f,l_f)=(0,0)\\
-\frac{2}{\pi r}(1-n)\sin\left(\frac{\pi n}{1-n}\right),~~~~~~~~~~~ \hskip 1cm\text{if}~ (n_f,l_f)=(n,1)\\
-\frac{2}{\pi r}(l_f-n_f)\sin\left(\frac{\pi n_f}{l_f-n_f}\right)-\frac{\cos{\theta}}{\pi}\left[2(1-l_f)-n+n_f\right]\sin\left[\frac{\pi(n-n_f)}{2(1-l_f)-n+n_f}\right]+\sin{\theta}\frac{n-n_f}{2},~\text{otherwise}.
\end{cases}
\end{equation}
The expression corresponding to the fully paired configuration with $n_f=l_f=0$ is nothing but Eq.~$\eqref{XX_GS_energy}$ enriched with the $U_1$ interaction energy density contribution arising from the formation of tight pairs only. Similarly, the energy density expression corresponding to the fully unpaired configuration with $n_f=1, l_f=0$ coincides with Eq.~$\eqref{gs_energy_perturbation}$: indeed, since the model aims at describing the onset of the P$_\pi$F-PS phase for large values of $r$, the kinetic energy contribution of unpaired fermions is estimated by restricting the ground state energy search in the subspace $\mathcal{H}_S$ of forbidden nearest-neighbor occupancy, where the interaction energy term proportional to $U_1$ evaluates to zero. Finally, the formula given for a properly phase-separated configuration is given by the sum of the two preceding expressions, after they have been straightforwardly generalized to the case where the unpaired fermions occupy an arbitrary fraction $l_f$ of the lattice.
 
Note that the parameters $n_f$ and $l_f$ are constrained by $n_f \in [0,n]$ and $l_f \in [2n_f, n_f+1-n]$. 
The latter range is an excluded-volume effect: $l_f\geq 2n_f$ because of forbidden nearest-neighbor occupancy on unpaired fermions; $n-n_f\leq 1-l_f$ because $k$ pairs occupy a region with at least $2k$ lattice sites.
\end{widetext}

\begin{figure}[b]
\centering
\includegraphics[width=\columnwidth,clip]{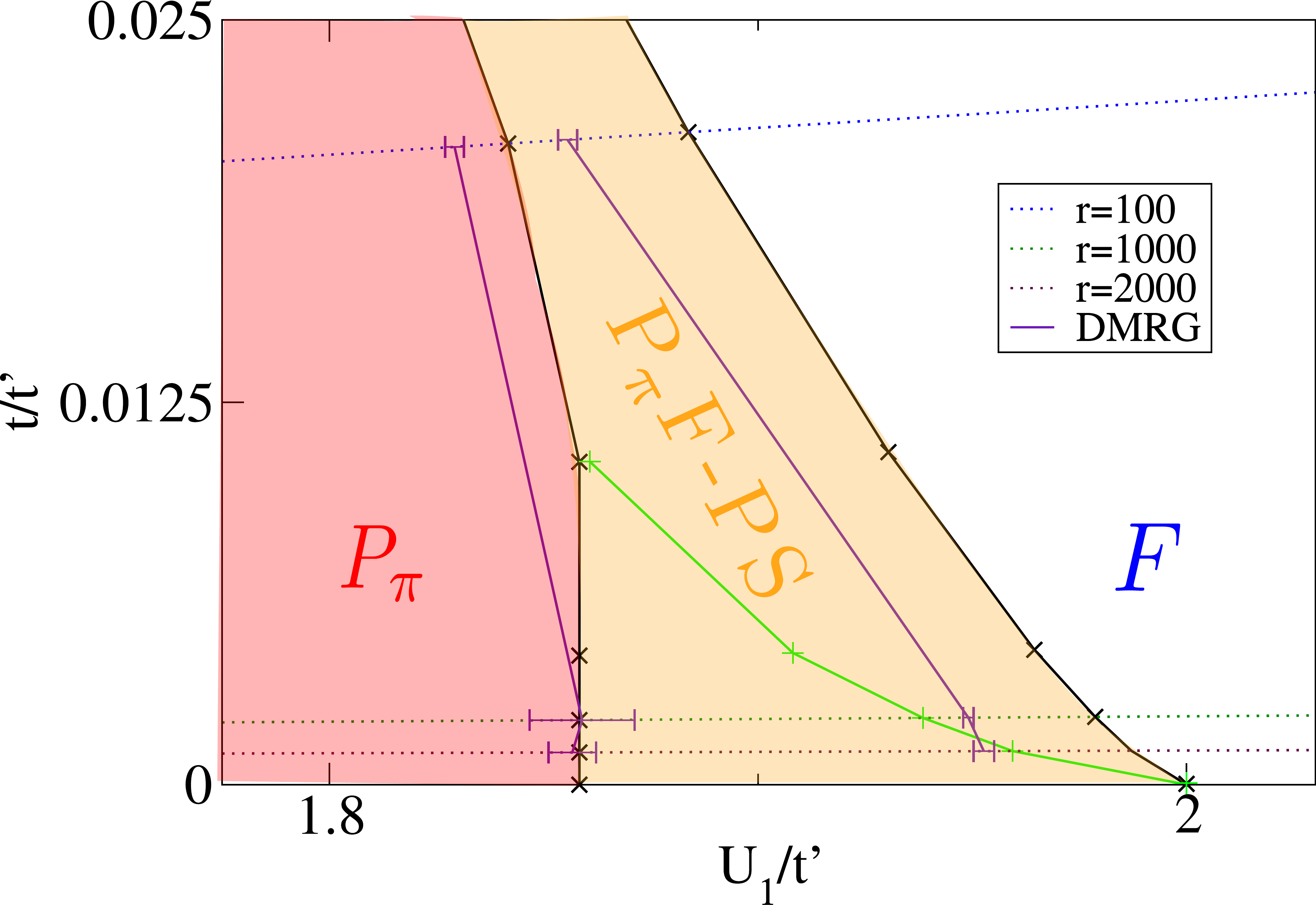}
\caption{Phase diagram in the $t/t'\to 0$ limit.
Black lines are the boundaries of the P$_\pi$F-PS phase obtained from model~\eqref{PS_model_general} for several radius $r$.
Violet lines are DMRG estimates of the phase boundaries for $r=100,1000,2000$. The green line sketches the boundary between the phase separation regime with a fermionic CDW domain (below the line) or with a fermionic liquid domain (above the line).}
\label{fig:PS_lines}
\end{figure}

The optimization of the energy ansatz $\mathcal E_2(n_f, l_f, r,\theta)$ in Eq.~\eqref{PS_model_general} leads to the phase diagram presented in Fig.~\ref{fig:PS_lines}, where we have plot the boundaries of the P$_\pi$F-PS phase (black lines) and compared them to the boundaries obtained for $r=100,1000,2000$ with DMRG simulations (violet lines). 
We find that the P$_\pi$F-PS phase is stable to the addition of a weak single particle hopping, which is in agreement with the DMRG numerical data. 
Although the diagram is given here for positive $t'$, the effective model is also valid for $t^{\prime}<0$ and a few DMRG results (not shown) support the existence of a P$_0$F-PS regime in this case too.

\subsubsection{Two different kinds of P$_\pi$F-PS phase separations}

As already mentioned in Fig.~\ref{fig:KE_PS_phase}, we find to kinds of P$_\pi$F-PS regimes, as the intrinsic features of the P$_\pi$F-PS phase evolve when $t/t'$ is increased. 
For $t=0$, the unpaired fermions are regularly arranged in a regular CDW pattern $|\cdots \bullet\circ \bullet \circ \cdots \rangle$. 
When $t \neq 0$, they either arrange in the aforementioned CDW crystal, or they delocalise and form a Luttinger liquid.
In the former case, the phase separation exhibits spatial segregation of a gapless (pairs) and of a gapped (fermions) phase; in the latter case, a spontaneous demixing of two gapless Luttinger liquid phases takes place.
The boundary between these two phase separation regimes is obtained with the effective model in Eq.~\eqref{PS_model_general} by locating the parameter values at which $n_f/l_f$ deviates from the CDW value of $1/2$. The result is shown in Fig.~\ref{fig:PS_lines} as a green line.

\begin{figure}[t]
\centering
\includegraphics[width=\columnwidth,clip]{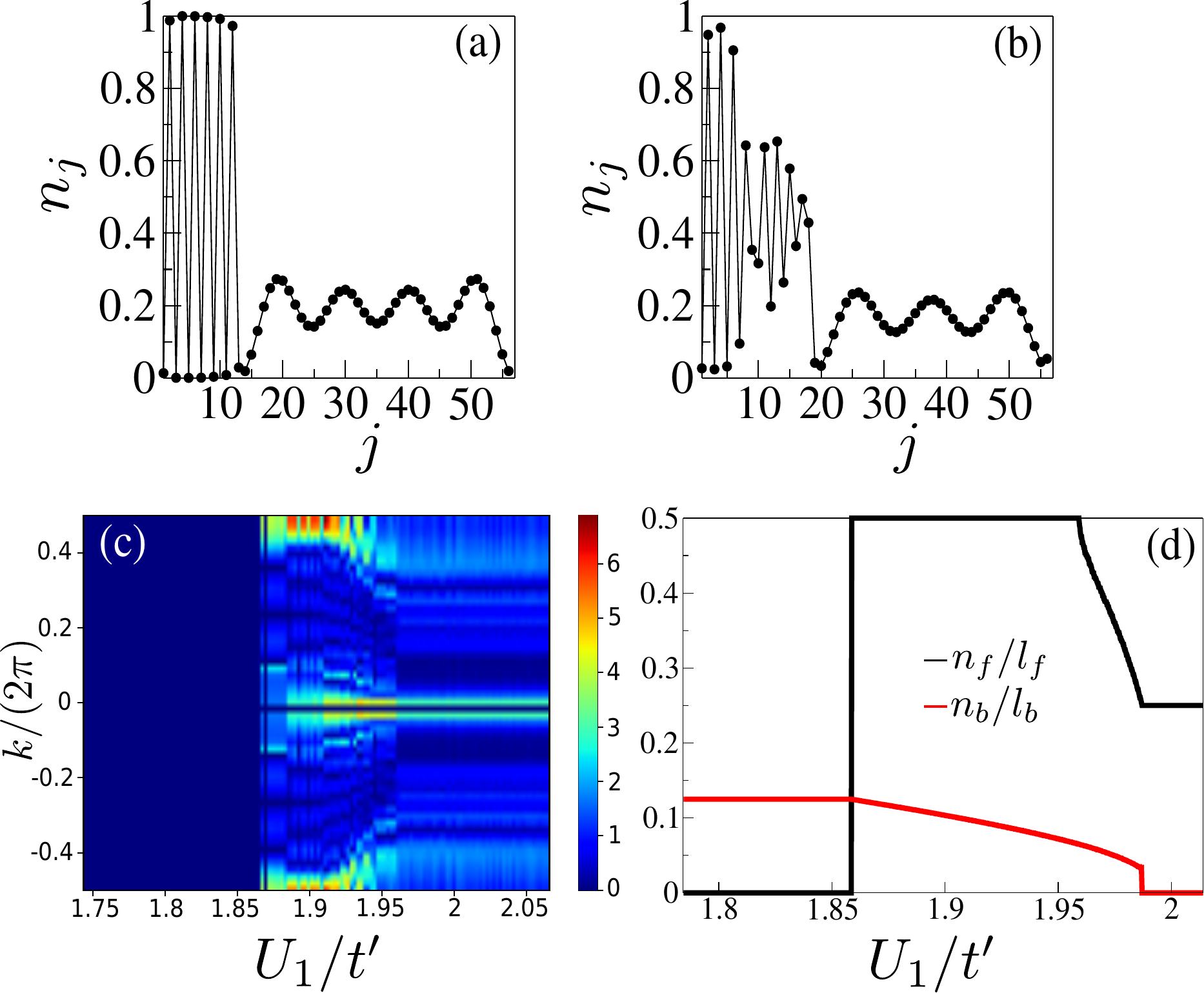}
\caption{Density profiles for the same parameters as in Fig.~\ref{fig:KE_PS_phase} ($r=2000$) (a) for a P$_\pi$F-PS phase with a CDW domain, (b) for a P$_\pi$F-PS phase with a liquid fermionic domain.
(c) Absolute value of $S(k)$ as a function of $U_1/t^{\prime}$.
(d) Predictions from the minimization of Eq.~\eqref{PS_model_general} for the fermionic density $n_f/l_f$ and the pair density $n_b/l_b$ as a function of $U_1/t^{\prime}$. }
\label{fig:N(k)}
\end{figure}

These qualitative differences are recovered in the numerical DMRG simulations performed at $r=2000$. This result strengthens the general predictive power of the effective model~\eqref{PS_model_general}. 
We provide direct evidence for the two forms of phase separation in Fig.~\ref{fig:N(k)}(a-b), where the local density profiles are shown. 
For the parameters of Fig.~\ref{fig:N(k)}(a), a clear CDW domain is observed, while in Fig.~\ref{fig:N(k)}(b), the local densities are compatible with a fluid domain (see also Fig.~\ref{fig:KE_PS_phase}(a-b) for the corresponding local kinetic energies).
In order to follow the evolution from a CDW to a fluid domain, we find it useful to compute the Fourier transform $S(k)$ of the density fluctuations in open boundary conditions:
\begin{equation}
 S(k) = \sum_j e^{-i k j} (\langle \hat n_j \rangle - n ).
\label{eq:Sk}
\end{equation}
This Fourier signal displays peaks corresponding to the main wavevectors of the fluctuations. For instance, the CDW pattern gives rise to a sharp peak at $k/(2\pi)=0.5$. The domains of pairs display a smooth peak at low-$k$ corresponding to the low pair density. On Fig.~\ref{fig:N(k)}(c), one can follow the both the transitions P$_\pi\to$P$_\pi$F-PS and P$_\pi$F-PS$\to$F (left to right) and, also, with the P$_\pi$F-PS phase, the transition from a CDW domain to a liquid one in the fermion region.
Note that such calculations are not easy numerically. One needs PBC  to remove edge effects, which are not optimal in DMRG and allow to address smaller systems than with OBC. Furthermore, since the domains are just a fraction of the total lattice, it is hard to reach large domain sizes. The $k$-discretization associated to small domain is clearly visible in Fig.~\ref{fig:N(k)}(c). Remarkably, these DMRG data are almost quantitatively captured by the effective model, as one can see in Fig.~\ref{fig:N(k)}(d). There, we plot the fluctuations peak expected from the local densities of the pair and fermionic domains. 
Such $k$-space patterns of the P$_\pi$F-PS phase are strongly different from those obtained in the C phase and which are recalled in Appendix~\ref{App:Struct:Factor:C}. Note that, due to the use of PBC, the P$_\pi$ phase has a perfectly constant density leading to $S(k)=0$ while, although one should expect the same for the F phase, a flat density profile is hard to achieve for unpaired fermions with such small values of $t$. This is why we observe a non-zero signal in the F phase. Yet, we observe that the profile is essentially unchanged in the F while increasing $U_1/t'$.

\subsection{Phase separation vs. coexistence phase}\label{Sec:Phase:Separation}

In the previous section, we have shown that in the large-$r$ limit the phase intervening between the F and the P$_\pi$ phases is characterised by phase separation. On the other hand, the phase diagram in Fig.~\ref{fig:phasediag} shows that for small $r$, the intermediate phase is the coexistence phase C.
It is thus natural to investigate in which way the C phase evolves into the P$_\pi$F-PS phase.

\begin{figure}
\centering
\includegraphics[width=\columnwidth,clip]{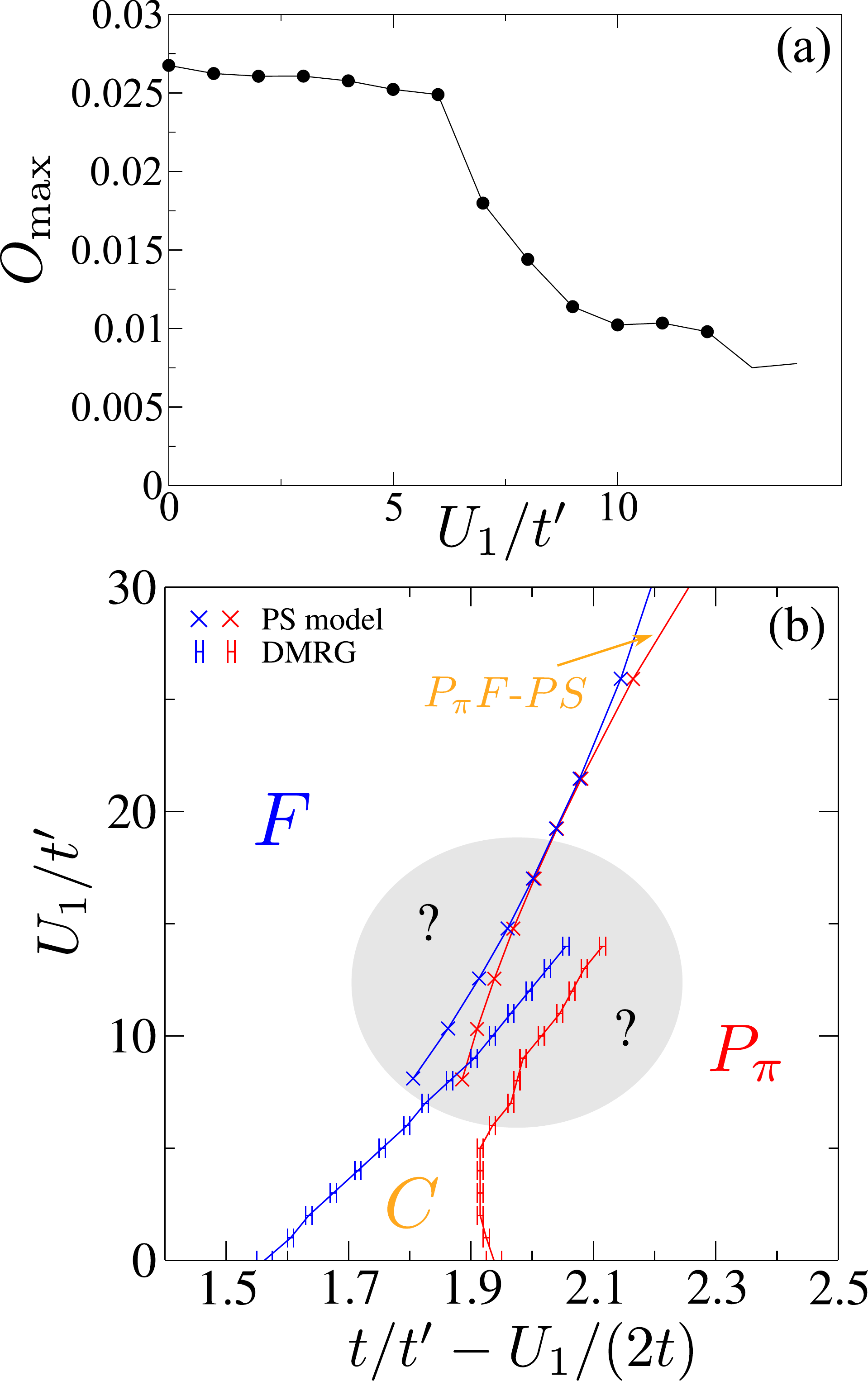}
\caption{(a) Maximum value $O_{\text{max}}$ along a cut at constant $U_1/t$ as a function of $U_1/t$. 
(b) Black lines indicate the boundaries of the P$_\pi$F-PS phase as predicted by model~\eqref{PS_model_general}. Red lines are DMRG estimates for the boundaries of the intermediate phase between F and P$_\pi$.}
\label{fig:overlaps}
\end{figure}

In order to discuss this point, we focus on the observable $O$ defined in Eq.~\eqref{Eq:Overlap}, which measures the overlap between the kinetic energy profiles of the unpaired fermions and of the pairs. We have already seen in Figs.~\ref{fig:KE_C_phase}(a) and Fig.~\ref{fig:KE_PS_phase}(c) that this quantity should be different from zero in the coexistence phase and equal to zero in the phase-separated phase. We now perform a systematic analysis of its behaviour for increasing radius $r$.
In Fig.~\ref{fig:overlaps}(a), we plot the maximum value of the overlap 
\begin{equation}
 O_{\text{max}}(U_1/t) = \max_{t'/t} \, O(t'/t, U_1/t)\,,
\end{equation}
along a cut at constant $U_1/t$, ranging from the F to the P$_\pi$ phase.
It measures the degree of spatial segregation of paired and unpaired fermions and therefore highlights qualitatively the crossover from one phase to the other. The maximum is always reached in the intermediate phase in between the F and the P$_\pi$ phase.
The finite size numerical data show that
$O_{\text{max}}$ remains stable with respect to its $U_{1}=0$ value up to $U_1/t=6$. 
This further supports the aforementioned stability of the C phase against nearest-neighbor density-density interactions. 
The value of $O_{\text{max}}$ drops for $U_1/t > 6$, which provides evidence for the progressive onset of phase separation observed at $t\ll t^{\prime},U_1$.

When $U_1/t \gtrsim 14$, the region separating the F and P$_\pi$ phases is shrinking so much that it makes the DMRG simulations particularly demanding. This shrinking is illustrated on the phase diagram of Fig.~\ref{fig:overlaps}(b), in which the abscissa has been rescaled to follow the main diagonal line of Fig.~\ref{fig:phasediag} along which this phase develops. The DMRG red lines are compared to the results of the effective large-$r$ model~\eqref{PS_model_general} pushed towards smaller values of $U_1/t$.
We observe that the latter model also predicts a shrinking of the P$_\pi$F-PS phase, until the two phase boundaries touch at $U_1/t \sim 20$. A possible scenario is that of a critical point taking place when the two boundaries touch which separates it from the C phase. Alternatively, the two phases might smoothly evolve one into the other, through a continuous crossover.
We leave as an open question the study of the onset of phase separation from the C phase and the determination of whether it occurs through a direct transition or a smooth crossover. Our results already provide essential guidelines for investigating this phase diagram.


%


\section{The two-fluid model}
\label{Sec:TwoFluid}

In this section we elaborate on the two-fluid model that we introduced in Ref.~[\onlinecite{Gotta2020}] and that was shown to provide a satisfactory description of the phase diagram for $U_1=0$ and $t'/t\geq0$.
The framework is mutuated from Ref.~[\onlinecite{Kane_2017}], where a phenomenological description of pairing in one-dimensional setups of spinless fermions was first proposed via the interaction of a bosonic fluid of pairs and a fermionic fluid of unpaired fermions.
We begin in Sec.~\ref{Sec:FermBoseDOF} by recalling the key properties of the model and by presenting some novel numerical data that motivate the description in terms of two fluids.
In Sec.~\ref{Sec:Rosch} we discuss an effective field-theory that incorporates quantum fluctuations into the two-fluid model and that describes satisfactorily the whole line $U_1=0$, and, with some minor modifications, the region $|U_1/t|\neq 0$ and small.

\subsection{Fermionic and Bosonic degrees of freedom} \label{Sec:FermBoseDOF}

The two-fluid model is based on the assumption that the system is populated by two species of particles, the unpaired fermions and the paired ones, which are hard-core bosons.
We introduce the Hamiltonians $\hat H_f$ and $\hat H_b$, which describe the fermionic and the bosonic species respectively:
\begin{subequations}
\label{Eq:Ham:2F:split}
\begin{align}
 \hat H_{f} =& -t \sum_j \hat d_j^\dagger \hat d_{j+1}+ H.c. \, ; \\
 \hat H_{b} =& + t' \sum_j \hat \sigma^+_j \hat \sigma^-_{j+1} + H.c. \,.
\end{align}
\end{subequations}
The  two species interact through a density constraint: $n = n_f + 2 n_b$, that is, through the fact that the bosons are composed of two fermions. As we discuss in the next section, other forms of interactions must be included in order to describe the whole phase diagram.

The success of the two-fluid model 
motivates the search for a microscopic numerical characterisation of the emerging degrees of freedom described by the $\hat d_j$ and $\hat \sigma_j^-$ operators appearing in the Hamiltonians~\eqref{Eq:Ham:2F:split}.
Indeed, they are two effective models and the fermionic $\hat d_j$ operators do not match the original fermions of the system $\hat c_j$;
similarly, the spin operators describe the pairs, and should not be simply seen as a product $\hat c_j \hat c_{j+1}$.
To better understand these statements, let us observe that the operator $\hat c_j$ detects also fermions that are paired; analogously the product $\hat c_j \hat c_{j+1}$ could detect two non-interacting fermions that are close by because of quantum delocalisation without any underlying pairing physics.

\begin{figure*}[t]
\centering
\includegraphics[width=0.9\textwidth]{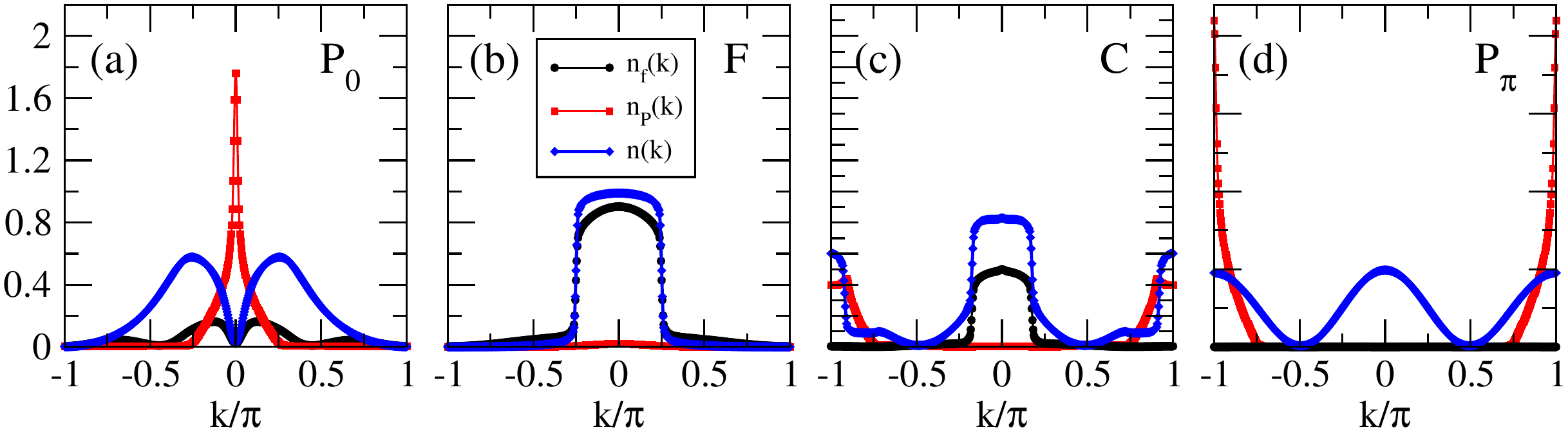} 
\caption{Effective occupation factors for unpaired fermions $n_f(k)$ and for pairs $n_P(k)$, see definitions in~\eqref{Eq:Eff:Occ:Factors}.  
The numerical parameters are $L=104$, $N=26,$ $U_1=0$. Different panels refer to different points of the phase diagram: (a) $t'/t = -2.2$ in the P$_0$ phase;
(b) $t'/t = 1.2$ in the F phase;
(c) $t'/t = 1.85$ in the C phase;
(d) $t'/t = 2.2$ in the P$_\pi$ phase. Simulations are performed with OBC.}
\label{fig:occ_factors_F}
\label{fig:occ_factors_P}
\label{fig:occ_factors_C}
\end{figure*}  

We introduce the following operators, which are expected to capture the main features of unpaired and paired fermions, respectively:
\begin{subequations}
\label{Eq:Truncated}
\begin{align}
 \hat f^{\dag}_j =& (1-\hat n_{j-1})\hat c^{\dag}_j (1-\hat n_{j+1}),\\
 \hat P^{\dag}_j =& (1-\hat n_{j-1})\hat c^{\dag}_j \hat c^{\dag}_{j+1}(1-\hat n_{j+2}),
 \end{align}
 \end{subequations}
The  projectors $1-\hat n_j$ play a key role because they enforce that the fermion $\hat f_j$ does not have any neighbor, and thus is truly unpaired. For the $\hat P_j$, the two projectors allow us to identify only paired fermions. 
In Appendix~\ref{App:H:pair:unpaires} we establish a mathematical link between the $\hat c_j$ operators and the two-fluid operators $\hat d_j$ and $\hat\sigma_j^-$; the operators~\eqref{Eq:Truncated} are a truncated version of the exact ones, and reproduce their properties with an accuracy that increases with the diluteness of the system.

We  study the effective occupation factors:
\begin{subequations}
\label{Eq:Eff:Occ:Factors}
\begin{align}
 n_{f}(k) = & \frac{1}{L}\sum_{j,l}e^{ik(j-l)}\langle \hat{f}^{\dag}_j \hat{f}_{l}\rangle; \label{F_occ_factor}\\
 n_{P}(k) = & \frac{1}{L}\sum_{j,l}e^{ik(j-l)}\langle \hat{P}^{\dag}_j \hat{P}_{l}\rangle. \label{P_occ_factor}
\end{align} 
\end{subequations}
The numerical results are presented in Fig.~\ref{fig:occ_factors_P}.
The F phase displays a fermionic occupation around $k \sim 0$; the P$_{0,\pi}$ phases display a quasi-condensate of pairs around the momenta $0$ and $\pi$, hence their names (see Ref.~[\onlinecite{Gotta2020}] for some effective models). On the other hand, the C phase displays the simultaneous presence of unpaired fermions around $k\sim 0$ and paired fermions around $k\sim\pi$.

As a concluding remark, it is worth mentioning that the standard occupation factor $n(k)=\frac{1}{L}\sum_{j,j^{\prime}}e^{ik(j-j^{\prime})}\langle \hat{c}^{\dag}_{j} \hat{c}_{j^{\prime}}\rangle$ displays different properties.
In Fig.~\ref{fig:occ_factors_P} we superimpose it to the curves $n_f(k)$ and $n_P(k)$. In the F phase, it is almost identical to $n_f(k)$, as expected. In the C phase, it displays a non-negligible occupation of momenta close to $k \sim 0$ and $k\sim \pi$, thus revealing its sensitivity both to unpaired fermions and to pairs. 
The sinusoidal form in the P$_\pi$ phase has been interpreted in Ref.~[\onlinecite{Gotta2020}] in terms of tightly-bound pairs whose size is not affected by the single-particle hopping, as it can be verified by simple perturbative arguments. 
The same reasoning is not valid in the P$_0$ phase: it shows instead that a perturbative value of $t$ increases the spatial extent of the pairs.
As a consequence, $n(k)$ is not sinusoidal. Moreover,
$n_f(k)$ does not vanish identically because the fermions constituting a single pair can be found at finite distance from each other in the ground state. 

\subsection{Quantum fluctuations and criticality in the two-fluid model} \label{Sec:Rosch}

The standard two-fluid model has been shown to describe the phases appearing for $t'/t>0$ enforcing only the constraint $n=n_f+2n_b$ and without any other direct form of interaction between the unpaired fermions and the pairs.
For $t'/t<0$, we observe a qualitatively different phase diagram which, as a consequence, must be shaped by interactions between the two species.
In the following, we show that
for $t'/t<0$ the interactions between the fluids correctly induce the critical point $c=3/2$ between F and P$_0$ phases.
The effective field theory that we are proposing thus provides a solid ground for describing the physics at $U_1=0$ and follows closely the treatments developed in Refs.~[\onlinecite{Kane_2017},\onlinecite{Sitte2009}].
Simple modifications allow to take into account the presence of interparticle interactions, $U_1 \neq 0$, for which the qualitative features of the phase diagram do not change.

We introduce the effective Hamiltonian $\hat H= \hat H_f+ \hat H_b+ \hat H_{int}$ where $\hat H_b$ is a Luttinger liquid Hamiltonian describing a partially-filled band of pairs and $\hat H_f$ describes a quadratic fermionic band of unpaired fermions:
\begin{subequations}
\label{eff_theory}
\begin{align}
\hat H_f=&\int dx\psi^{\dag}\left(\epsilon_0-\frac{\partial_{x}^{2}}{2m}-\mu\right)\psi,\\
\hat H_b=&\frac{v}{2\pi}\int dx\left[K(\partial_x\phi)^{2}+\frac{1}{K}(\partial_{x}\theta)^{2}-2\mu\left(\frac{\partial_x\theta}{2\pi}+\rho_b\right)\right],
\end{align}
\end{subequations}
where the chemical potential $\mu$ couples to the density of paired and unpaired fermions and $[\phi(x),\theta(x^{\prime})]=i\pi\Theta(x-x^{\prime})$ and $\rho_b$ is the average pair density.
The Hamiltonian $H_{int}$ describes the interaction between unpaired fermions and pairs, which takes the form of a process that transforms two fermions into a pair (and viceversa):
\begin{equation}
H_{int}=\int dx\,u(x)\left(e^{i2\phi}\psi\partial_{x}\psi+H.c.\right). \label{Eq:ux}
\end{equation}

This field theory should thus contains all the relevant information for describing the system when the fermionic mode is unoccupied or starts to be occupied.
The key quantity is the function $u(x)$ that appears in Eq.~\eqref{Eq:ux}.
When $t^{\prime}/t<0$, the coupling is non-oscillating, $u(x)=u$, because the pairs and the fermions are located around $k \sim 0$ and during the exchange process there is no momentum transfer.
In this limit, the model has been studied in Ref.~[\onlinecite{Kane_2017}] where it is shown that a bosonic phase with a gap to single particle excitations and a fermionic phase coinciding with a standard Luttinger liquid are separated by a critical point with central charge $c=3/2$.

On the other hand, when $t^{\prime}/t>0$, the paired states lie in reciprocal space around $\pm\pi$, thus implying that whenever two fermions get converted into a pair, a net momentum transfer of $\sim \pi$ takes place. The function $u(x)$ inherits this spatial modulation, $u(x) \sim e^{i \pi x/a} \sim (-1)^{x/a}$ and is then averaged out by the integral. This process requires a momentum exchange and thus is strongly suppressed,
while the next leading contribution involves the conversion of four fermions into two pairs and can be neglected. The resulting phenomenology has been studied in Ref.~[\onlinecite{Balents2000}] and is known to exhibit a Lifschitz transition to a coexistence phase with a pair of gapless modes, consistently with the behaviour found in such a parameter regime.

We conclude this section by noting that an alternative approach for dealing with pairing phenomena in charge-conserving spinless systems exists, and it is based on the use of an emergent mode~\cite{Ruhman_2017, he_emergent_2019}.
In the appendices~\ref{Sec:Non:Interacting}, \ref{App:Bosonisation:NonInteracting}, \ref{App:Bosonization:correlators} and~\ref{App:Bosonization:adhoc_term} we discuss a bosonisation approach to our model and present some considerations on how the use of an emergent mode could describe the physics of the coexistence phase (for the $c=3/2$ transition see Ref.~[\onlinecite{Ruhman_2017}]).

\section{Conclusions}\label{Sec:Conclusions}

In this article we have discussed the phase diagram of the RA model \cite{Ruhman_2017}, which proved extremely rich, and have presented numerical simulations for all encountered phases. 

Specifically, we have focused on two questions. The first one is related to the understanding of the difference of the coexistence phase \cite{Gotta2020} from a phase-separated phase. 
A phase separation occurs in various parts of the phase diagram and we have indeed shown that the coexistence phase can evolve into a phase-separated phase. It is important to stress that this transition takes place at non-perturbative values of the interaction, so that the stability of the coexistence phase is well established. The precise characterization of the nature of the transition from coexistence to phase separation is left for future work.

The second question we have addressed is related to the discussion of pairing physics in one-dimensional systems of spinless fermions, which is known to elude standard bosonisation approaches and to require the ad-hoc introduction of either an emergent mode \cite{Ruhman_2017, he_emergent_2019} or of a two-fluid model \cite{Kane_2017, Gotta2020}. This article demonstrates that employing a many-fluid approach can be extremely fruitful when applied to pairing physics, since it can describe ample parts of the phase diagram with a simple and intuitive picture. This work opens the path to further studies and to the application of this many-fluid approach to other situations where pairing or generalization thereof have been invoked for the appearance of symmetry-enriched Majorana fermions or non-local parafermions~\cite{Chew_2018, Mazza_2018, Calzona_2018}.

\acknowledgements

We thank E.~Orignac and J.~Ruhman for enlightening discussions. We are grateful to A.~Kantian and I.~Mahyaeh for a discussion on flat bands. 
We acknowledge funding by LabEx PALM (ANR-10-LABX-0039-PALM).  
This work has been supported by Region Ile-de-France in the framework of the DIM Sirteq.

\appendix 

\section{Structure factor in the coexistence phase}
\label{App:Struct:Factor:C}

\begin{figure}[h]
 \includegraphics[width=\columnwidth,clip]{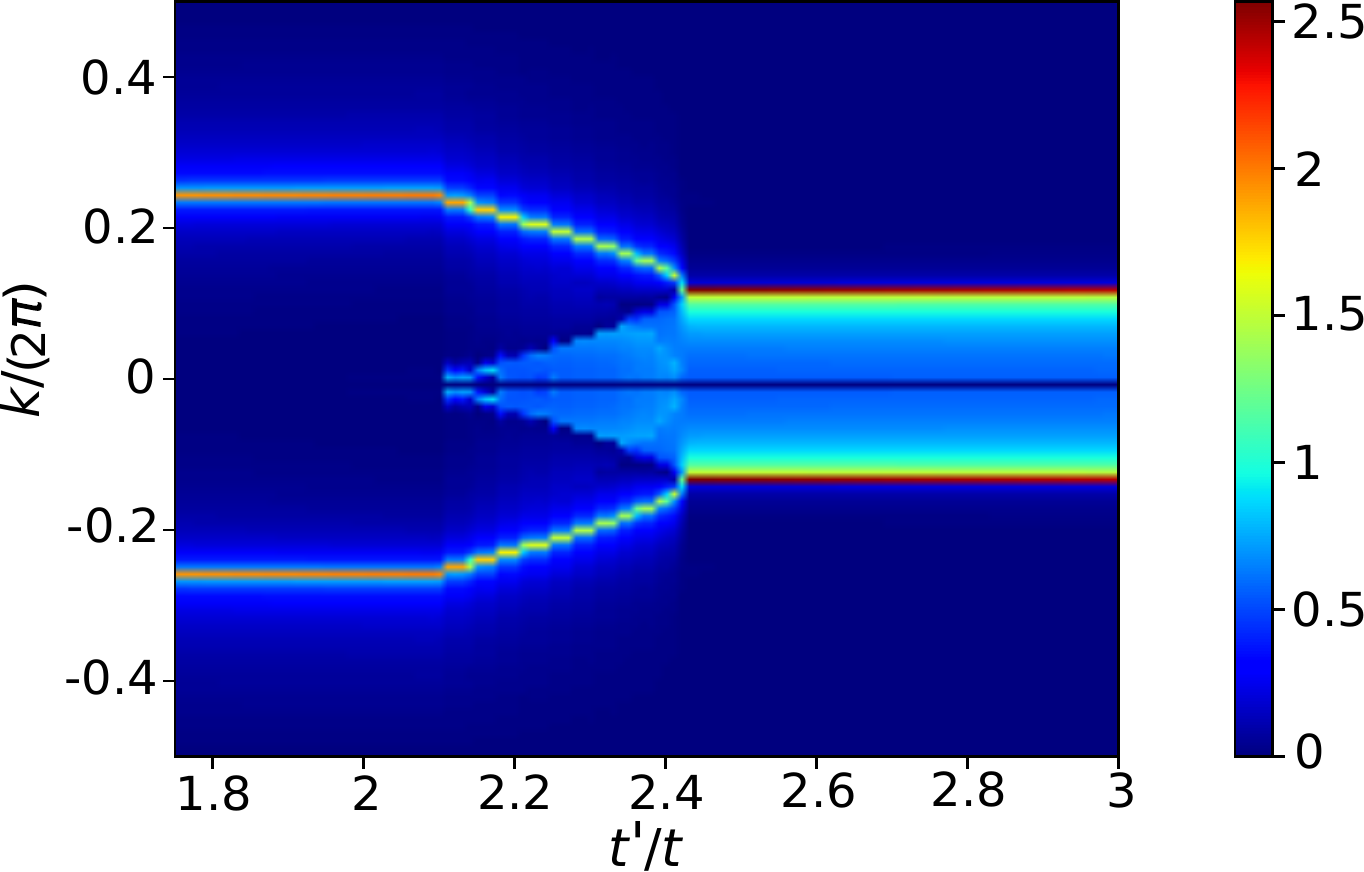}
 \caption{Map of the absolute value of the Fourier transform $S(k)$ as a function of $t^{\prime}/t$ for $U_1/t=1$ and L=104 in OBC.}
 \label{fig:N(k)_C_phase}
\end{figure}

We briefly recall the behavior of the Fourier transform of the density fluctuations~\ref{eq:Sk} for the coexistence phase C in Fig.~\ref{fig:N(k)_C_phase}, computed with OBC and to be compared with Fig.~\ref{fig:N(k)}(c) that was computed with PBC.
The robustness of the features of the C phase against the $U_1$ term is striking in the behavior of the peaks in the Fourier transform of the density fluctuations shown in Fig.~\ref{fig:N(k)_C_phase}. The F and P$_\pi$ phases are characterized, respectively, by peaks at $k=2\pi n$ and $k=2\pi (n/2)$, as expected from a standard Luttinger liquid and a Luttinger liquid of pairs. In the intermediate C phase, the behavior of the peaks has been interpreted in Ref.~\cite{Gotta2020} as a signature of a mixture pairs and unpaired fermions fluids.
By denoting the effective densities of unpaired fermions as $n_f$ and the effective pair density as $n_b$, the resulting uncoventional peaks are to be found at $k=2\pi(n_f+n_b)$, $k=2\pi n_b$ and $k=2\pi n_f$, the latter being a subleading contribution.

\section{Description of the Hilbert space in terms of paired and unpaired fermions}\label{App:H:pair:unpaires}

Ideally, it would be desirable to find a set of creation and annihilation operators associated to the paired and unpaired degrees of freedom in the system and obtain an explicit formulation of the Hamiltonian in terms of them. To proceed in this this direction, we start by observing that fermions in a generic Fock position basis state will occur in $l$-particle clusters: if $l$ is even, we interpret it as $l/2$ contiguous pairs; if $l$ is odd, we view the cluster as $(l-1)/2$ consecutive pairs followed by an unpaired fermions. As an example, let us consider the state $\ket{\bullet\bullet\bullet\circ\bullet\bullet\circ}=\hat{c}^{\dag}_1\hat{c}^{\dag}_2\hat{c}^{\dag}_3\hat{c}^{\dag}_5\hat{c}^{\dag}_6\ket{0}$; then, the first cluster of size $3$ is reinterpreted as a pair followed by an unpaired fermion, while the second cluster of size $2$ is considered as a pair.

\begin{widetext}
After achieving a description of the Hilbert space in terms of paired and unpaired fermions, we are now able to write down the mathematical expression of their corresponding creation operators. In particular, a pair is created on sites $j,j+1$ if and only if the number of fermions preceding those sites is even, whereas an unpaired fermion is created at site $j$ if and only if it is preceded by an even number of fermions and followed by none.  More precisely, the expressions for the unpaired fermion creation operator $\hat{f}^{\dag}_j$ and the paired fermions creation operator $\hat{P}^{\dag}_j$ read: 
\begin{subequations}\label{App:pair:unpaires:operators}
\begin{align}
&\hat{f}^{\dag}_j=\left[1-\hat{n}_{j-1}+(1-\hat{n}_{j-2l-1})\sum_{l\geq1}\left(\prod_{m=1}^{2l}\hat{n}_{j-m}\right)\right]\hat{c}^{\dag}_j\left(1-\hat{n}_{j+1}\right)\\
&\hat{P}^{\dag}_j=\left[1-\hat{n}_{j-1}+(1-\hat{n}_{j-2l-1})\sum_{l\geq1}\left(\prod_{m=1}^{2l}\hat{n}_{j-m}\right)\right]\hat{c}^{\dag}_j\hat{c}^{\dag}_{j+1}.
\end{align}
\end{subequations}
The expressions~\eqref{App:pair:unpaires:operators} imply that the creation and annihilation operators for these effective degrees of freedom have a nonlocal character, as the bare lattice creation and annihilation operators get dressed by projectors with increasingly large support. Referencing the concrete example presented above, the definition of the operators $\hat{f}^{\dag}_j$ and $\hat{P}^{\dag}_j$ implies in particular that $\ket{\bullet\bullet\bullet\circ\bullet\bullet\circ}=\hat{P}^{\dag}_1\hat{f}^{\dag}_3\hat{P}^{\dag}_5\ket{0}$.
\end{widetext}

As a result, it is more convenient to resort to numerical estimates of the appropriately truncated versions of the above nonlocal operators. Indeed, in the dilute limit, the expression for the operators $\hat{f}^{\dag}_j$ and $\hat{P}^{\dag}_j$ can be truncated by keeping only the first term in parenthesis, obtaining as a result the operators defined in~\eqref{Eq:Truncated} and concretely analyzed in numerical simulations. As we are working in the dilute limit, the addition of the factor $1-\hat{n}_{j+2}$ to the right of the term $\hat{c}^{\dag}_j\hat{c}^{\dag}_{j+1}$ in the numerically studied observable is not expected to modify qualitatively the results.

\section{Bosonization for $t' \ll t$}
\label{App:Bosonisation:NonInteracting}

In this section, we develop a bosonization treatment of the system around the free-fermion point in the weak coupling regime $t^{\prime}\ll t$. Following standard recipes~\cite{giamarchi_quantum_2010}, we expand the lattice operators around two Fermi points $\pm k_F$ in terms of two long wavelength fermionic field operators $\psi_R(x),\psi_L(x)$ as $c_j\sim \sqrt{a}\left[\psi_R(ja)e^{ik_F ja}+\psi_L(ja) e^{-ik_F ja}\right]$, $a$ being the lattice spacing. 
By re-expressing the fermionic fields in terms of a canonically conjugated pair of bosonic fields $\phi(x), \partial_x\theta(x)$ satisfying $\left[\phi(x),\partial_{x^{\prime}}\theta(x^{\prime})\right]=i\delta(x-x^{\prime})$, it is possible to rewrite the Hamiltonian as an effective bosonic field theory describing the low energy physics of the system.

We rewrite the number operator appearing in the pair kinetic energy as $n_j=n+:n_j:$, which amounts to explicitly decouple the average and fluctuation contributions to the local density. Therefore, the effective low energy Hamiltonian takes the quadratic form:
\begin{equation}
H=\frac{v}{2}\int dx \left[K\left(\partial_x\theta\right)^{2}+\frac{1}{K}\left(\partial_x\phi\right)^{2}\right],
\end{equation}
where $v$ is the velocity of the acoustic mode and $K$ is the Luttinger parameter, given by:
\begin{equation}
K=\frac{1}{\sqrt{1-\frac{4t'\cos\left(2k_F a\right)}{\pi\left[ t \sin(k_F a)+2nt'\sin(2k_F a)\right]}}}.
\end{equation}

The repulsive $U_1$ term in the Hamiltonian can also be included in the treatment, as it amounts to the quadratic term $\frac{U_1 a}{\pi}\left[1-\cos(2k_F a)\right]\int dx (\partial_x \phi)^{2}$ and simply contributes to a renormalization of the effective parameters $u$ and $K$.

As expected, the weak-coupling bosonization approach predicts a single bosonic mode for all parameter values and hence is inadequate to  capture the transition neither  the coexistence phase nor to any of the P$_{0,\pi}$ phases. Indeed, the latter implies a nonperturbative reshaping of the ground state Fermi surface associated to the emergence of the gapless excitation mode associated to the liquid of pairs. The single mode Luttinger liquid approach does not allow to explore such a phenomenology and thus requires to be complemented by a more refined treatment. This is what we do next.

\section{A bosonisation description of the $U_1=0$ line}
\label{Sec:Non:Interacting}

In this section we present an analytical approach to the description of the properties of the phase diagram along the line $U_1=0$ that is based on
a bosonisation model 
that takes as a starting point the paired phases.

\subsubsection{An unconventional non-interacting starting point}

In order to make progress towards the low-energy description of the C phase in bosonization language, we partition the lattice operators $\hat c_j$ into two groups, corresponding to even and to odd sites. Introducing the notation $c_{j,1}=c_{2j},c_{j,2}=c_{2j-1}$, the Hamiltonian 
can be rewritten as:
\begin{align}
\hat H=&-t\sum_j\left[\hat c^{\dag}_{j,1}\left(\hat c_{j,2}+ \hat c_{j+1,2}\right)+H.c.\right]+ \nonumber \\
&-t^{\prime}\sum_j \left[\hat n_{j,1}\hat c^{\dag}_{j,2}\hat c_{j+1,2}+\hat  n_{j,2}\hat c^{\dag}_{j-1,1}\hat c_{j,1}+H.c.\right],
\end{align} 
and it can be interpreted as the model of a zig-zag ladder. We rewrite the operators $\hat n_{j,\alpha}$ as the sum of the average density  and of the density fluctuations via the relation $\hat n_{j,\alpha}=n+:~\hat n_{j,\alpha}:$, the Hamiltonian becomes the sum of a quadratic term:
\begin{align} \label{H0_hamiltonian}
\hat H_{0}=&-t\sum_j\left[\hat c^{\dag}_{j,1}\left(\hat c_{j,2}+\hat c_{j+1,2}\right)+H.c.\right]+\nonumber \\
&- n t^{\prime}\sum_{j}\left[\hat c^{\dag}_{j,2}\hat c_{j+1,2}+\hat c^{\dag}_{j-1,1}\hat c_{j,1}+H.c.\right]
\end{align} 
and of a quartic term $\hat V$ that we do not need to specify for the moment.

\begin{figure}[t]
\centering
\includegraphics[width=\columnwidth,clip]{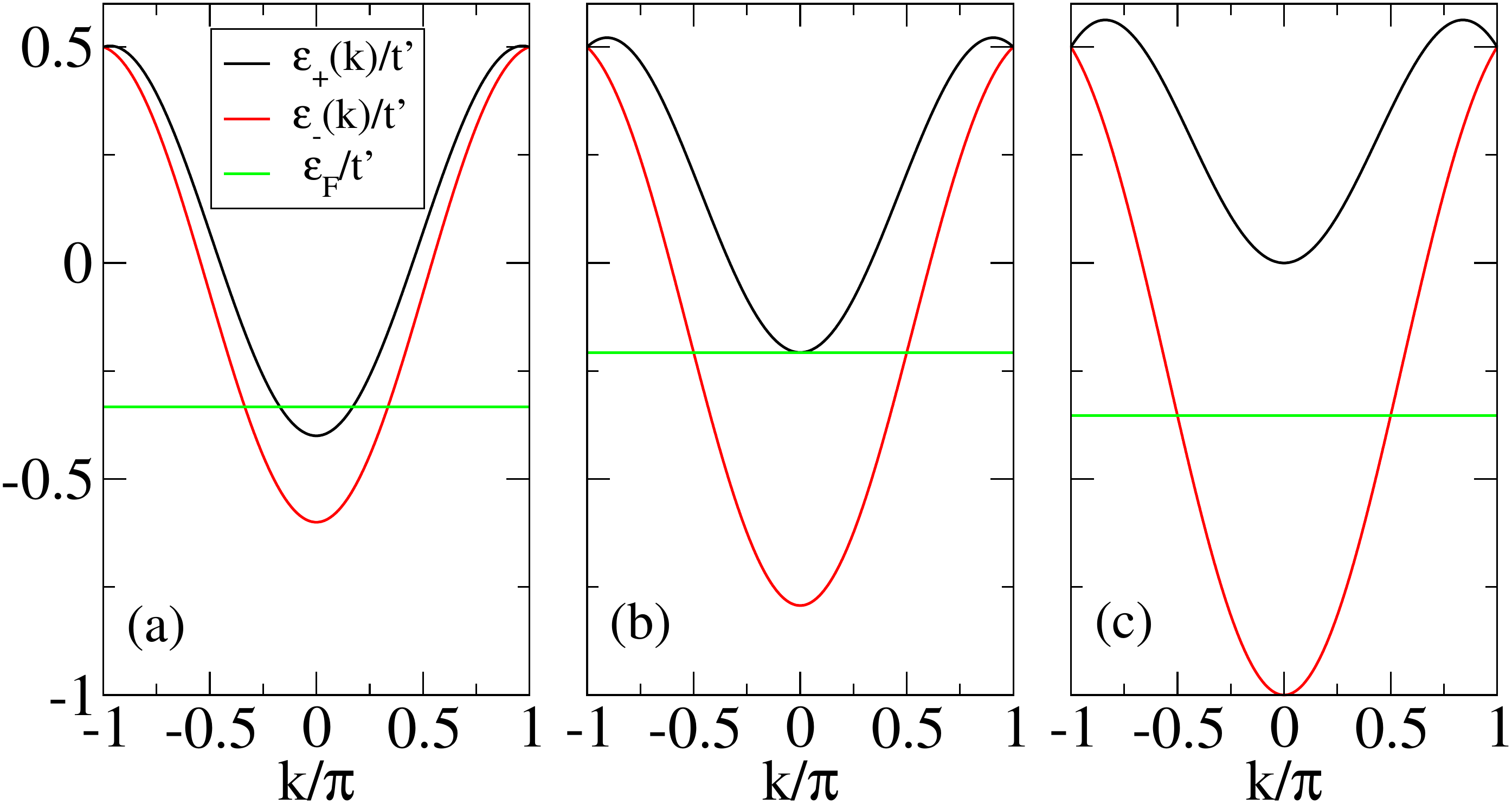}
\caption{Dispersion relations (\ref{dispersion_relations}) of the quadratic Hamiltonian $\hat H_{0}$ in Eq.~\eqref{H0_hamiltonian} and corresponding Fermi energy $\varepsilon_{F}$ for $ n=0.25$ and (a) $t'/t=20$, (b) $t'/t=2\sqrt{2}(1+\sqrt{2})$ and (c) $t'/t=4$.  }
\label{fig:disp_rel}
\end{figure}

The hamiltonian $\hat H_{0}$ can be diagonalized 
with a Fourier transform followed by a Bogoliubov rotation and can be written as
\begin{subequations}
\begin{align}
\hat H_{0}=&\sum_{k}\sum_{\alpha=\pm}\epsilon_{\alpha}(k)  \hat \gamma^\dagger_{k,\alpha} \hat \gamma_{k,\alpha}; \\
\epsilon_{\pm}(k)=&-2nt^{\prime}\cos(k)\pm 2 t \cos\left(\frac{k}{2}\right),\label{dispersion_relations}
\end{align}
\end{subequations}
where the operators $\hat \gamma_k$ are related to the original $\hat c_j$ operators.
As shown in Fig.~\ref{fig:disp_rel}, the above dispersion relations introduce an additional pair of Fermi points at $n=1/4$ when $t'/t > 2 \sqrt{2} (1+ \sqrt 2) $,
thus realizing a more promising starting point for the derivation of an effective low-energy field theory that captures the properties of the C phase. 
The above value of $t'/t$ is obtained by asking that $\epsilon_+(0) = \epsilon_-(2\pi n)$ for $t, t'>0$ and in general it yields:
\begin{equation}
 \frac{t'}{t} =  \frac{1}{n}\frac{1+\cos(\pi n)}{1- \cos(2 \pi n)}.
\end{equation}

It is important to observe that the dispersion relations that we have found are identical to those presented in Ref.~[\onlinecite{Ruhman_2017}], apart from a folding into a halved Brillouin zone due to the rewriting of the Hamiltonian in terms of two coupled chains with an effective doubled lattice spacing.
In fact, the bosonisation approach that we are going to describe can be regarded as an extension of the method presented in Ref.~[\onlinecite{Ruhman_2017}] to the case of $t'/t>0$.

\subsubsection{Bosonisation description of interactions}

We introduce a pair of canonically-conjugated bosonic field operators $\phi_{\alpha}(x)$ and $\partial_{x}\theta_{\alpha}(x)$ for the two pairs of gapless points $\pm k_{F,\alpha}$ appearing on the curves $\epsilon_{\alpha}(k)$ in~\eqref{dispersion_relations}. 
The non-interacting Hamiltonian becomes $\hat H_{0}=\sum_{\alpha} \frac{v_{F,\alpha}}{2}\int dx\left[(\partial_{x}\phi_{\alpha})^{2}+(\partial_{x}\theta_{\alpha})^{2} \right]$, where $v_{F,\alpha}$ is the Fermi velocity of the band $\alpha$.

Concerning the interaction term 
\begin{align}
\hat V=&-t^{\prime}\sum_j \biggl[:\hat n_{j,1}:\left(\hat c^{\dag}_{j,2}\hat c_{j+1,2}+H.c.\right)+ \nonumber \\
& \qquad \qquad \quad  +:\hat n_{j,2}:\left(\hat c^{\dag}_{j-1,1}\hat c_{j,1}+H.c.\right)\biggl],
\end{align}
its bosonised form is derived in an effective way considering all allowed momentum-conserving two-particle processes around the Fermi points.
We first consider density-density forward scattering, which produces a term $\sum_{\alpha,\beta}g_{\alpha\beta}\int dx \frac{1}{\pi}\partial_{x}\phi_{\alpha}\partial_x \phi_{\beta}$ and that eventually only renormalises the Luttinger parameter of the bosonic field Hamiltonian $\hat H_0$.
An additional family of momentum conserving processes is represented by the terms transferring particles between the two bands,
proportional to $\int dx \left(\psi^{\dag}_{R,+}\psi^{\dag}_{L,+}\psi_{R,-}\psi_{L,-}+H.c.\right)$, which translates to $-g_1\int dx \cos\left[2\sqrt{\pi}(\theta_{+}-\theta_{-})\right]$.

After performing the canonical transformation  $\phi_{0}=\phi_{+}+\phi_{-}$, 
$\theta_{0}=\theta_{-}$, 
$\phi_{1}=\phi_{+}$, 
$\theta_{1}=\theta_{+}-\theta_{-}$, the Hamiltonian takes the general form:
\begin{align} 
H_{bos}=&\sum_{\alpha=0,1}\frac{v_{\alpha}}{2}\int dx\left[K_{\alpha}(\partial_x \theta_{\alpha})^{2}+\frac{1}{K_{\alpha}}(\partial_{x}\phi_{\alpha})^{2}\right]+\nonumber \\
&+l_{1}\int dx\partial_x \phi_{1}\partial_x \phi_0 +l_2 \int dx \partial_x \theta_{1}\partial_x \theta_0+\nonumber \\
 &-g_{1}\int dx \cos\left[2\sqrt{\pi}\theta_1\right].
 \label{Eq:Field:Bosonisation:1}
 \end{align}
The terms proportional to $l_1$ and $l_2$ are a consequence of the final canonical transformation and do not alter the nature of the low energy excitations; the effective Luttinger-liquid parameters $v_{0,1},K_{0,1}$ 
have been introduced to take into account non-linearities. 

When $g_1$ is relevant, the field $\theta_1$ is pinned to a minimum of the cosine term and the model describes a paired phase. 
Indeed, in Appendix~\ref{App:Bosonization:correlators} we show that this phase features quasi-long-range order in pair correlations and exponentially decaying single-particle correlation functions, due to the gap in the single-particle spectrum, as it was originally shown in Ref.~[\onlinecite{Ruhman_2017}].
The question now is to understand whether the inclusion of physically-motivated perturbations to~\eqref{Eq:Field:Bosonisation:1} allows us to describe (i) the direct $c=3/2$ transition to the F phase, and (ii) the transition to the F phase separated by the C phase.

Concerning point (i),
in Ref.~[\onlinecite{Ruhman_2017}], it was shown how to capture the critical behavior of the model for $t^{\prime}/t<0$: 
The idea is to add to the low-energy field theory~\eqref{Eq:Field:Bosonisation:1} the most relevant and general term giving a mass to the emergent mode, i.e., a term proportional to $\int dx \left[\left(\psi^{\dag}_{R,+}\psi_{L,+} \right)+H.c.\right]\propto \int dx \cos(2\sqrt{\pi}\phi_1)$. 
The resulting field theory implies a direct transition from a standard Luttinger liquid phase, corresponding to the pinning of the field $\phi_1$, to a Luttinger liquid of pairs, when the pinned field is $\theta_1$, where the single-particle excitations have become gapped, while the pair excitations remain gapless (see details in Appendix~\ref{App:Bosonization:correlators}). The critical point separating the two phases belongs to the Ising universality class and is characterised by a  central charge $c=3/2$, as it has been numerically verified by fitting the entanglement entropy profile of the ground state~\cite{Ruhman_2017}.

Concerning point (ii), in order to have a non-direct transition to the F phase, as it happens for $t'/t>0$, we find that we need to introduce terms that describe the interaction between the two bands at higher order.
In particular, we consider this $m$-particle term proportional to $\int dx \left[\left(\psi^{\dag}_{R,+}\psi_{L,+} \right)^{m}+H.c.\right]$, which has the following bosonised form: 
\begin{equation}
\hat H_m = -\tilde g_m \int dx \cos(2m\sqrt{\pi}\phi_1) .
\end{equation}
In Appendix~\ref{App:Bosonization:adhoc_term}, we show that when $m>2$,
the model $H_{bos}+ H_m$ features three different phases as a function of $K_1$. For low values of $K_1$, $\tilde g_m$ is relevant and $g_1$ is irrelevant: with arguments similar to those mentioned above, this phase can be identified with the F phase.
For large values of $K_1$, $\tilde g_m$ is irrelevant and $g_1$ is relevant: as above, this phase is identified with a paired phase.
The novelty is constituted by the fact that there is an intermediate phase for intermediate values of $K_1$, where both $\tilde g_m$ and $g_1$ are irrelevant. This phase is characterized by a central charge $c=2$ and it is tempting to identify it with the C phase. However, at this stage, we have not been able to produce a direct bosonisation characterisation of this phase that allows a more precise identification beyond the central charge. We interpret this phase as a mean-field precursor of the C phase adiabatically connected to the latter. Indeed, the pairing term has been approximated by an average-density-assisted hopping term in constructing the unconventional starting point for the treatment of interactions, thus trading a genuine pairing term with single-particle processes.

We conclude with a speculation concerning the difference between the P$_0$ and P$_\pi$ phases. If this analysis is correct, it implies that the difference between the two phases consists in the way interactions are included in bosonisation. For the P$_0$ phase, one can directly include $\hat H_{m=1}$ while for the P$_\pi$ phase, one needs $\hat H_{m>2}$.

\section{Behavior of the correlators}
\label{App:Bosonization:correlators}

We show that the behavior of the single particle correlator predicted by the effective field theory in 
Eq.~\eqref{Eq:Field:Bosonisation:1} is consistent with the phases appearing in the phase diagram of the model for positive values of $t^{\prime}/t$. First of all, it can be shown from the diagonalization of the Hamiltonian in Eq.~\eqref{H0_hamiltonian} that the lattice operators $c_{j,\alpha},\,\alpha=1,2$ can be expressed as:
\begin{subequations}
\begin{align}
&\hat{c}_{j,1}=\frac{1}{\sqrt{2}}(\hat{c}_{j+\frac{1}{4},-}-i\hat{c}_{j+\frac{1}{4},+})\\
&\hat{c}_{j,2}=\frac{1}{\sqrt{2}}(\hat{c}_{j-\frac{1}{4},-}+i\hat{c}_{j-\frac{1}{4},+}),
\end{align}
\end{subequations}
where $\hat{c}_{j\pm\frac{1}{4},\beta}=\frac{1}{\sqrt{\frac{L}{2}}}\sum_k e^{ik(j\pm\frac{1}{4})}\hat{c}_{k,\beta},\,\beta=\pm$, indicating as $L$ the length of the starting lattice. 
Expressing the lattice operators $\hat{c}_{j,\pm}$ in terms of the long-wavelength bosonic fields describing excitations around the Fermi points of the dispersion relation in Eq. \eqref{dispersion_relations}, one obtains  for example:
\begin{widetext}
\begin{align}
\hat{c}_{j,1}=\frac{1}{\sqrt{4\pi\alpha}}\biggl[&e^{ik_{F,-}(x+\frac{a}{4}})e^{-i\sqrt{\pi}[\theta_0 (x+\frac{a}{4})-\phi_0(x+\frac{a}{4})+\phi_1(x+\frac{a}{4})]}+e^{-ik_{F,-}(x+\frac{a}{4})}e^{-i\sqrt{\pi}[\theta_{0}(x+\frac{a}{4})+\phi_0(x+\frac{a}{4})-\phi_1(x+\frac{a}{4})]}+ \nonumber \\
&-ie^{ik_{F,+}(x+\frac{a}{4})}e^{-i\sqrt{\pi}[\theta_0(x+\frac{a}{4})+\theta_1(x+\frac{a}{4})-\phi_1(x+\frac{a}{4})]}
-ie^{-ik_{F,+}(x+\frac{a}{4})}e^{-i\sqrt{\pi}[\theta_0(x+\frac{a}{4})+\theta_1(x+\frac{a}{4})+\phi_1(x+\frac{a}{4})]} \biggr].
\label{operator_fields}
\end{align}
\end{widetext}
Here, $\alpha$ is a regularization cutoff and $a$ is the lattice spacing along each one of the two subchains that the original lattice has been divided into and thus equals twice the lattice spacing between sites in the latter. An analogous expression holds for $\hat{c}_{j,2}$.

When the field $\theta_{1}$ is pinned, the field $\phi_1$ is completely disordered and has exponentially decaying correlations. As the latter appears in each term of the expression of $\hat{c}_{j,1}$ and $\hat{c}_{j,2}$, every single particle correlator of the form $G_{\alpha,\alpha^{\prime}}(r)=\langle \hat{c}^{\dag}_{j,\alpha}\hat{c}_{j+r,\alpha^{\prime}}\rangle$ with $\alpha=1,2$ decays exponentially with distance. Therefore, it is natural to interpret this phase as a pairing phase with a gap to single particle excitations. In particular, in the main text we  interpreted the phase characterized by the pinning of the field $\theta_1$ as the $P_\pi$ phase.

On the other hand, when the field $\phi_1$ is pinned, the field $\theta_1$ is disordered. As a result, the field $\phi_1$ can be replaced by its expectation value. The last two terms in the equation~\eqref{operator_fields} contain explicitly the field $\theta_1$, allowing one to disregard their contribution. As a result, the leading contribution to the lattice operators is given by the first two terms of Eq.~\eqref{operator_fields} (as in a single mode Luttinger liquid), thus certifying that the leading contribution to $G_{\alpha,\alpha^{\prime}}(r)$ is algebraic in $r$, as expected for a standard Luttinger liquid at weak coupling. 
In the main text we argue that the inclusion of a term proportional to $\int dx \cos\left(2\sqrt{\pi}n\phi_1\right)$ allows to reintroduce the F phase to the low energy description~\eqref{Eq:Field:Bosonisation:1}.  


\section{Phenomenological inclusion of the F phase}
\label{App:Bosonization:adhoc_term}
Let us consider a general field theory of the form:
\begin{subequations}
\begin{align}
&H=H_{LL}(\phi_{0},\theta_{0};v_{0},K_{0})+H_{LL}(\phi_{1},\theta_{1};v_{1},K_{1})\nonumber\\
&-g_1\int dx\cos(\beta_1\theta_{1})-g_2\int dx\cos(\beta_2\phi_{1}),
\end{align}
\end{subequations}
where $H_{LL}(\phi,\theta;v,K)=\frac{v}{2\pi}\int dx\left[K\left(\partial_x\theta\right)^{2}+\frac{1}{K}\left(\partial_{x}\phi\right)^{2}\right]$ and the fields satisfy $[\phi_{\alpha}(x),\partial_{x^{\prime}}\theta_{\alpha^{\prime}}(x')]=i\pi\delta_{\alpha,\alpha^{\prime}}\delta(x-x^{\prime})$. The first order RG equations for the couplings read:
\begin{align}
&\frac{dg_1}{dl}=\left(2-\frac{\beta_{1}^{2}}{4K_{1}}\right)g_1\\
&\frac{dg_2}{dl}=\left(2-\frac{\beta_{2}^{2}}{4}K_{1}\right)g_2,
\end{align}
which implies that the simultaneous irrelevance condition for the cosine terms takes the form:
\begin{align}
& 2-\frac{\beta_{1}^{2}}{4K_{1}}<0\\
& 2-\frac{\beta_{2}^{2}}{4}K_{1}<0.
\end{align}
which in turn means that, for both cosines to be irrelevant, the Luttinger parameter must satisfy $\frac{8}{\beta_{2}^{2}}<K_1<\frac{\beta_{1}^{2}}{8}$. In order for such a parameter regime to exist, one has to impose $\frac{8}{\beta_{2}^{2}}<\frac{\beta_{1}^{2}}{8}$, which forces $\beta_1\beta_2>8$, assuming $\beta_1$ and $\beta_2$ to be positive, as the cosine is an even function.\\
After the rescaling $\theta^{\prime}_{\alpha}=\sqrt{\pi}\theta_{\alpha}, \phi^{\prime}_{\alpha}=\sqrt{\pi}\phi_{\alpha}$ (leading to $[\phi^{\prime}_{\alpha}(x),\partial_{x^{\prime}}\theta_{\alpha^{\prime}}^{\prime}(x')]=i\pi\delta_{\alpha,\alpha^{\prime}}\delta(x-x^{\prime})$), the effective theory (\ref{Eq:Field:Bosonisation:1}) has the general structure:
\begin{align}
&H=H_{LL}(\phi^{\prime}_{0},\theta^{\prime}_{0};v_{0},K_{0})+H_{LL}(\phi^{\prime}_{1},\theta^{\prime}_{1};v_{1},K_{1})\\
&-g_1\int dx\cos(2\theta^{\prime}_{1}),
\end{align}
i.e., $\beta_{1}=2$. If we generalize the class of mass terms which gap the emergent mode to  general $n^{th}$-order processes, we need to consider terms of the form:
\begin{equation}
\int dx\left[\left(\psi^{\dag}_{R,+}\psi_{L,+}\right)^{n}+H.c.\right]\sim\int dx\cos\left(2n\phi^{\prime}_{1}\right),
\end{equation}
which means that $\beta_{2}=2n$. As a result, the condition $\beta_1\beta_2>8$ turns into $n>2$, i.e., one needs to add phenomenologically processes of order $3$ or higher as the most relevant contribution to the Hamiltonian such that the weak coupling Luttinger liquid phase is reintroduced to the theory and, at the same time, a $c=2$ coexistence phase is stabilized.

\bibliography{paper.bib}

\end{document}